\RequirePackage{lineno} 

\documentclass[jkps,preprint,fleqn,showkeys,superscriptaddress]{revtex4}
\usepackage{graphicx}
\usepackage{amssymb}
\usepackage{amsmath}
\usepackage{caption}
\usepackage{subcaption}
\usepackage{multirow}
\usepackage{bm}
\usepackage{xcolor}
\usepackage{hyperref}

\usepackage{xspace}

\usepackage{orcidlink}

\newcommand{\orcid}[1]{\href{https://orcid.org/#1}{\textcolor[HTML]{A6CE39}{\aiOrcid}}}

\begin{document}
\setcounter{page}{0}
\title[]{Identification of $tqg$ flavor-changing neutral current interactions using machine learning techniques}

\author{Byeonghak Ko \orcidlink{0000-0001-9974-4453}}
\author{Jeewon Heo \orcidlink{0000-0003-4463-4104}}
\author{Woojin Jang \orcidlink{0000-0002-1571-9072}}
\author{Jason S. H. Lee \orcidlink{0000-0002-2153-1519}}
\author{Youn Jung Roh \orcidlink{0009-0002-9335-9903}}
\email{uosyoun14@uos.ac.kr}
\author{Ian James Watson \orcidlink{0000-0003-2141-3413}}
\affiliation{Department of Physics, University of Seoul, Seoul 02504, Republic of Korea}
\author{Seungjin Yang \orcidlink{0000-0001-6905-6553}}
\affiliation{Department of Physics, Kyung Hee University, Seoul 02453, Republic of Korea}

\newcommand{\unit}[1]{\textrm{ #1}}
\newcommand{\SaJa}{\textsc{SaJa}}

\newcommand{\MpT}{\vec{p}_T^{\textrm{miss}}}
\newcommand{\MET}{E_T^{\textrm{miss}}}

\newcommand{\Br}{{Br}}

\newcommand{\plotsizeFull}{10cm}
\newcommand{\plotsizeHalf}{5.5cm}

\newcommand{\fig}[1]{\textrm{Figure~\ref{#1}}}
\newcommand{\tbl}[1]{\textrm{Table~\ref{#1}}}

\newcommand*{\met}{\ensuremath{p_{T}^{\;\textrm{miss}}}\xspace}
\newcommand*{\vecmet}{\ensuremath{\vec{p}_{T}^{\;\textrm{miss}}}\xspace}

\newcommand*{\cqg}{\ensuremath{C_{qg}}\xspace}
\newcommand*{\cug}{\ensuremath{C_{ug}}\xspace}
\newcommand*{\ccg}{\ensuremath{C_{cg}}\xspace}

\newcommand*{\cqgsqr}{\ensuremath{C_{qg}^{\;\;2}}\xspace}
\newcommand*{\cugsqr}{\ensuremath{C_{ug}^{\;\;2}}\xspace}
\newcommand*{\ccgsqr}{\ensuremath{C_{cg}^{\;\;2}}\xspace}

\begin{abstract}
Flavor-changing neutral currents (FCNCs) are forbidden at tree level in the Standard Model (SM), but they can be enhanced in physics Beyond the Standard Model (BSM) scenarios.
In this paper, we investigate the effectiveness of deep learning techniques to enhance the sensitivity of current and future collider experiments to the production of a top quark and an associated parton through the $tqg$ FCNC process, which originates from the $tug$ and $tcg$ vertices.
The $tqg$ FCNC events can be produced with a top quark and either an associated gluon or quark, while SM only has events with a top quark and an associated quark. 
We apply machine learning techniques to distinguish the $tqg$ FCNC events from the SM backgrounds, including $qg$-discrimination variables.
We use the Boosted Decision Tree (BDT) method as a baseline classifier, assuming that the leading jet originates from the associated parton.
We compare with a Transformer-based deep learning method known as the Self-Attention for Jet-parton Assignment (\SaJa) network, which allows us to include information from all jets in the event, regardless of their number, eliminating the necessity to match the associated parton to the leading jet.
The \SaJa\ network with qg-discrimination variables has the best performance, giving expected upper limits on the branching ratios $\Br(t \to qg)$ that are 25--35\% lower than those from the BDT method.

\end{abstract}

\keywords{FCNC, top quark, Transformer-based, deep learning, self-attention, machine learning}

\maketitle

\section{Introduction}
\label{sec:Intro}

Flavor-changing neutral currents (FCNCs) in the Standard Model (SM) are forbidden at the tree level and suppressed at higher orders through the Glashow–Iliopoulos–Maiani mechanism \cite{PhysRevD70GIM}.
At the one-loop level of the SM, top quark FCNC branching ratios $\Br(t \to ug)$ (respectively, $\Br(t \to cg)$) are of the order of $10^{-14}$ ($10^{-12}$) \cite{ActaPhysPolonB04AS}.
However, FCNC interactions may enhance these branching ratios to up to $10^{-4}$ in scenarios including physics Beyond the Standard Model (BSM), such as the $Q=2/3$ quark singlet model \cite{PhysRevD03ASQS}, the two Higgs doublet model \cite{PhysRevLett99AguilaTHDM}, and the minimal supersymmetric Standard Model \cite{NulPhysB99GuashMSSM}.

\begin{figure}[b]
    \centering
    \subfloat[][Diagrams with the $tg$ production]{\label{subfig:FeynmanDiagramFCNCUG}
      \includegraphics[width=14.5cm]{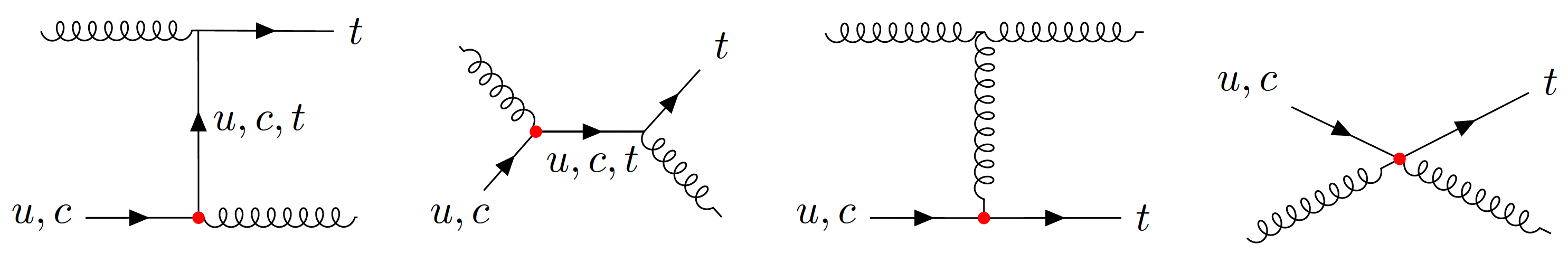}
    }
    
    \subfloat[][Diagrams with $tq$ productions]{\label{subfig:FeynmanDiagramFCNCOtherA}
      \includegraphics[width=11.5cm]{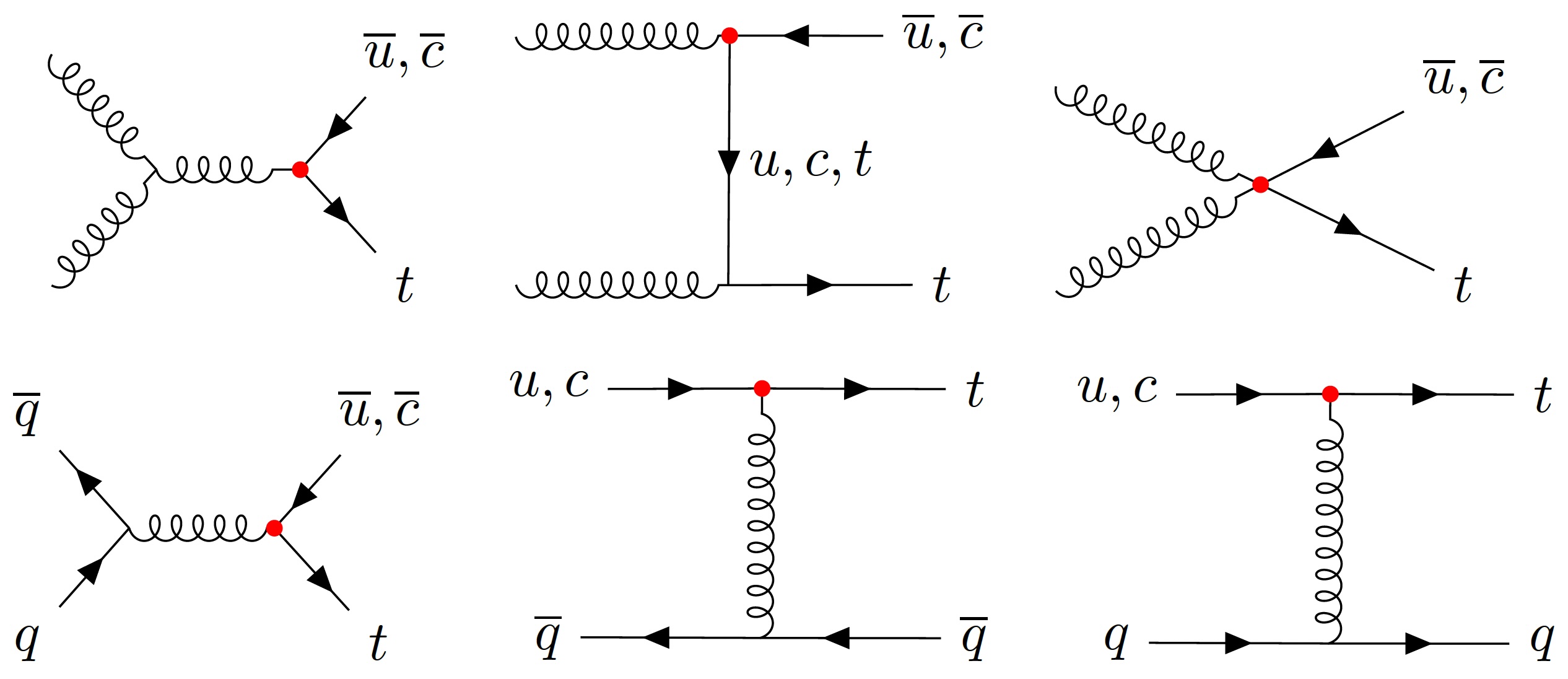}
    }
    \caption{The tree-level Feynman diagrams depicting the $tqg$ FCNCs. The vertices that are colored red indicate the $tug$ or $tcg$ FCNC vertices in the diagram.}
    \label{fig:FeynmanDiagramFCNC}
\end{figure}

In this study, we focus on $tqg$ FCNCs, where $q$ is $u$ or $c$, in proton--proton ($pp$) collisions at a center-of-mass energy of $13 \unit{TeV}$ to probe BSM physics.
The tree-level Feynman diagrams for the $tqg$ FCNCs are shown in \fig{fig:FeynmanDiagramFCNC}, where the red dots indicate the $tug$ and $tcg$ vertices for $tg$ production (\fig{subfig:FeynmanDiagramFCNCUG}) and $tq$ production (\fig{subfig:FeynmanDiagramFCNCOtherA}).
The $tq$ production resembles SM events, while the $tg$ production is forbidden at the tree level in the SM due to the conservation of charge and quark numbers.
However, the proportion of $tg$ production is significantly higher than that of $tq$ production, enabling us to distinguish $tqg$ FCNC events from the SM.

Searches for $tqg$ FCNCs have been performed by the ATLAS \cite{PhysLettB12ATLAStqg,EurPhysJC16ATLAStqg,EurPhysJC22ATLAStgq} and CMS \cite{JHEP17CMStgq} Collaborations in $pp$ collisions at center-of-mass energies of 7, 8, and 13 TeV, and there is also a feasibility study to search for the $tqg$ FCNC interactions in future colliders \cite{PhysRevD19Oyulmaz}.
These studies used machine learning techniques with high-level features as inputs to produce their final results.
However, they did not utilize gluon discrimination to enhance their sensitivity to the $tqg$ FCNC interactions.
In this study, we employ jet-based $qg$-discrimination variables \cite{CMSQGDiscPAS,CMSQGDiscProceeding,QGDiscSYCNN,QGDiscSYWSL} to distinguish the $tqg$ FCNC events from the SM backgrounds.

We utilize the boosted decision trees (BDTs) as a baseline method for event classification~\cite{BDT}.
We assign the leading jet as the jet originated from the associated parton, but the matching efficiency of the leading jet to the associated parton is approximately 60\%.
To improve over the baseline method by taking advantage of the full event topology, we adopt the Transformer-based deep learning method called the Self-Attention for Jet-parton Assignment (\SaJa) network \cite{SaJa}.
\SaJa\ is able to utilize the information from all the jets in the event, regardless of the number of jets, thereby eliminating the need to match the associated parton to the leading jet.

\section{Simulation}
\label{sec:TheoryAndSim}

To simulate the signal events, we use the \textsc{TopFCNC} model \cite{TopFCNCUFO}, which is an implementation of the $tqg$ FCNC Lagrangian in the \textsc{FeynRules} package \cite{FeynRules} in the universal \textsc{FeynRules} output format \cite{UFO}.
The effective Lagrangian terms involving the $tqg$ FCNCs can be written as follows:
\begin{eqnarray}
  \frac{\cqg}{\Lambda^2} g_s \overline{t} \sigma^{\mu \nu} T^a q \tilde{\varphi} G^a_{\mu \nu} + \textrm{h.c.}, \;\; \label{eq:EFTFCNCMain} \nonumber
\end{eqnarray}
where $\Lambda$ is the scale of the new physics, 
$g_s$ is the coupling constant of the strong interaction, 
$T^a$ are the generators of the $SU(3)$ gauge group, 
$G^a_{\mu \nu}$ is the gluon field strength tensor, 
and \cqg ($q = u, c$) is the strength of the $tqg$ FCNC interaction.
We set $\cqg / \Lambda^2$ as $0.2 \unit{TeV}^2$ for both $q = u, c$.

The $tqg$ FCNC events in $pp$ collision at 13 TeV are simulated using \textsc{MadGraph 5} v2.6.7 \cite{MG5} at the leading order (LO) with the 5-flavor scheme and the NNPDF 3.1 parton distribution function \cite{NNPDF31,LHAPDF6}.
The parton shower is performed using \textsc{Pythia} v8.224 \cite{Pythia8} with Monash 2013 tune \cite{TuneMonash2013}.
The events contain at most two additional partons in the hard process, using the MLM scheme to merge the events~\cite{MLM}.
We have observed that approximately 80\% (50\%) of events from $tug$ ($tcg$) vertex have $tg$ production.
Detector simulation is carried out using \textsc{Delphes} v3.4.2 \cite{Delphes} with the default CMS detector configuration card that comes with \textsc{Delphes}.

We generate SM background events with the same simulation setup as the $tqg$ FCNC signal samples.
The samples we generate are the single top quark t-channel process, top quark pair production, and $W$+jets.
Like FCNC processes, single top quark t-channel events have at most two additional partons in the hard process.
Other potential backgrounds, such as single top quark processes associated with $W$ bosons, Drell-Yan, and multijet backgrounds, are disregarded due to their comparatively negligible contributions after event selection.

\section{Event selection}
\label{sec:EvtSel}

The event topology of the $tqg$ FCNC events is similar to the single top quark t-channel process.
Therefore, we follow an event selection based on a recent CMS study, using Run II data, of the single top quark t-channel process, where the top quark decays leptonically~\cite{EurPhysJC20CMSTchDiff}.
We select muons and electrons with $p_T > 30 \unit{GeV}$, $|\eta| < 2.4$, and isolation $< 0.06$, using a relative particle flow isolation with $\Delta R < 0.4 \; (0.3)$ for muons (electrons)~\cite{PFIso}.
We discard events with any additional lepton with $p_T > 15 \unit{GeV}$, $|\eta| < 2.4$, and isolation $< 0.20$.

Jets are reconstructed from the particle flow outputs using the anti-$k_T$ algorithm \cite{antikT} with a cone size of $\Delta R = 0.4$ as implemented in \textsc{FastJet} v3.3.2 \cite{FastJet}.
We select jets with $p_T > 40 \unit{GeV}$ and $|\eta| < 2.4$.
We reject jets where the $\Delta R$ between the jet and the selected lepton is smaller than $0.4$.
We simulate $b$-jet tagging, following the $b$-tagging efficiency and mistag rates from the CMS CSVv2 medium working point \cite{CMSCSVv2,CMSCSVMore}, by updating the parameterized $b$-jet tagging efficiency and mistag rates in the \textsc{Delphes} card.
We select events with at least two jets and require that events have at least one $b$-tagged jet.

We define the transverse $W$ mass as the transverse mass of the reconstructed lepton and the reconstructed missing transverse momentum, \vecmet.
We reject events where the transverse $W$ mass is smaller than $50 \unit{GeV}$ to suppress multijet background events.

We reconstruct the top quark from the selected lepton, $b$-jet, and $\vecmet$ using a method based on the previous $tqg$ FCNC search by CMS~\cite{JHEP17CMStgq}.
We assume that $\vecmet$ corresponds to the neutrino $\vec{p}_T$ and we estimate the component $p_z$ of the neutrino momentum along the beam direction by requiring the lepton and neutrino give the $W$ boson mass $M_W = 80.4 \unit{GeV}$ \cite{PDGTable}, which leads to a quadratic equation whose solution is $p_z$.
In the case of real solutions, the smallest value is taken as $p_z$.
If the solutions are complex, we modify \vecmet to make the transverse $W$ mass equal to $M_W$ to eliminate the imaginary part of the solution.
The modified \vecmet is used only for the top quark reconstruction.

\section{Analysis strategy}
\label{sec:Strategy}

\begin{figure}
  \includegraphics[width=\plotsizeHalf]{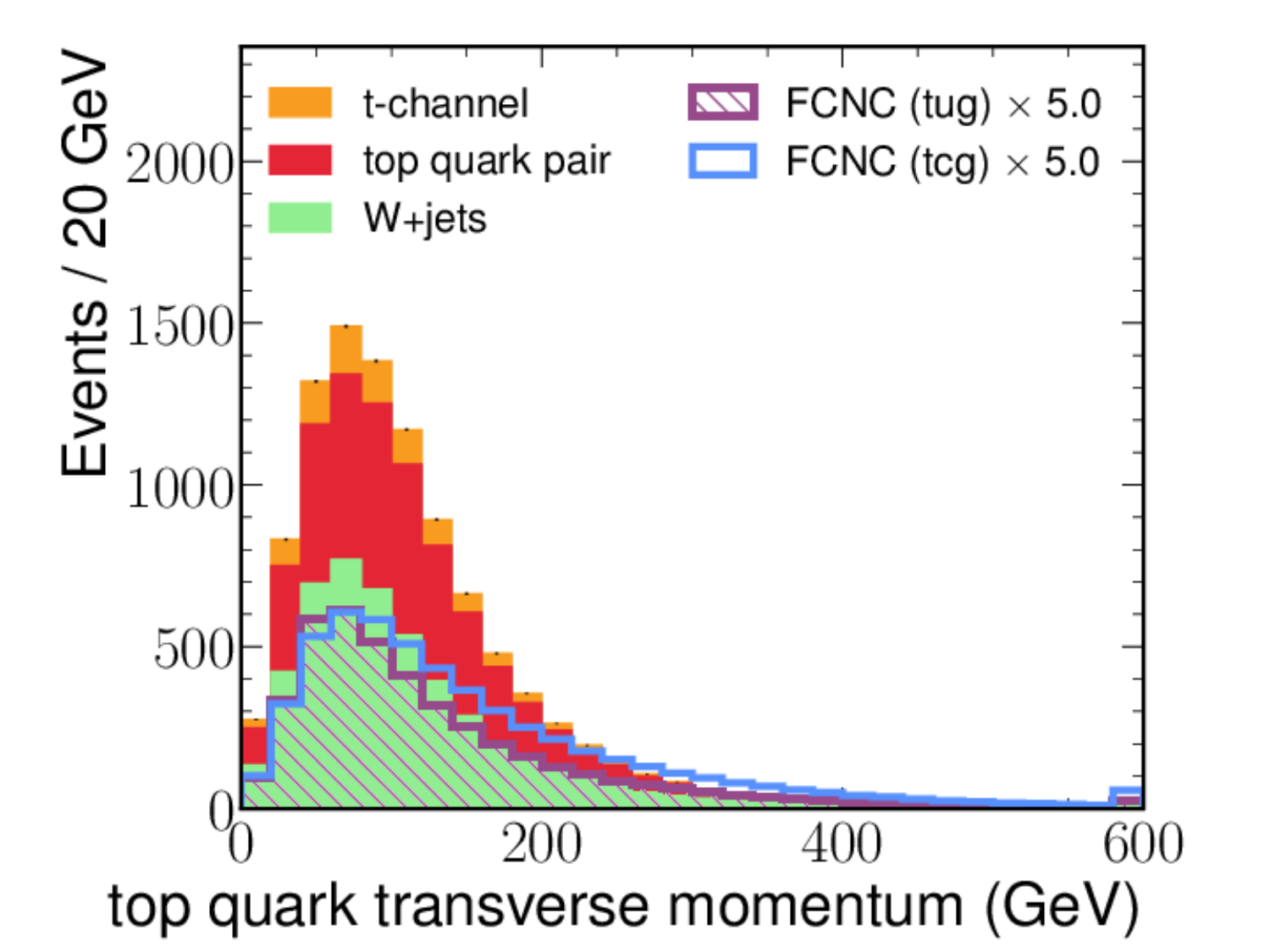}
  \includegraphics[width=\plotsizeHalf]{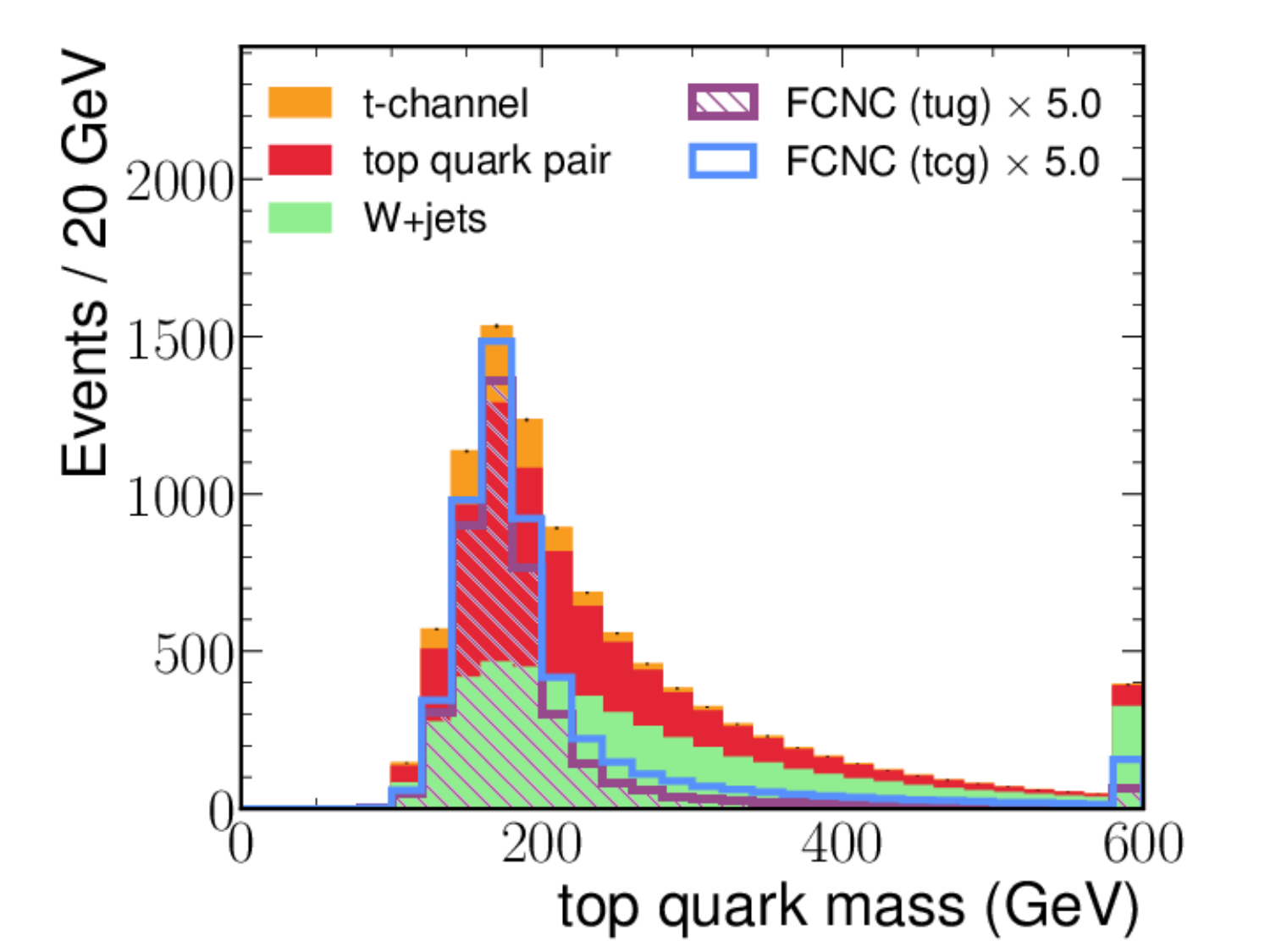}
  
  \includegraphics[width=\plotsizeHalf]{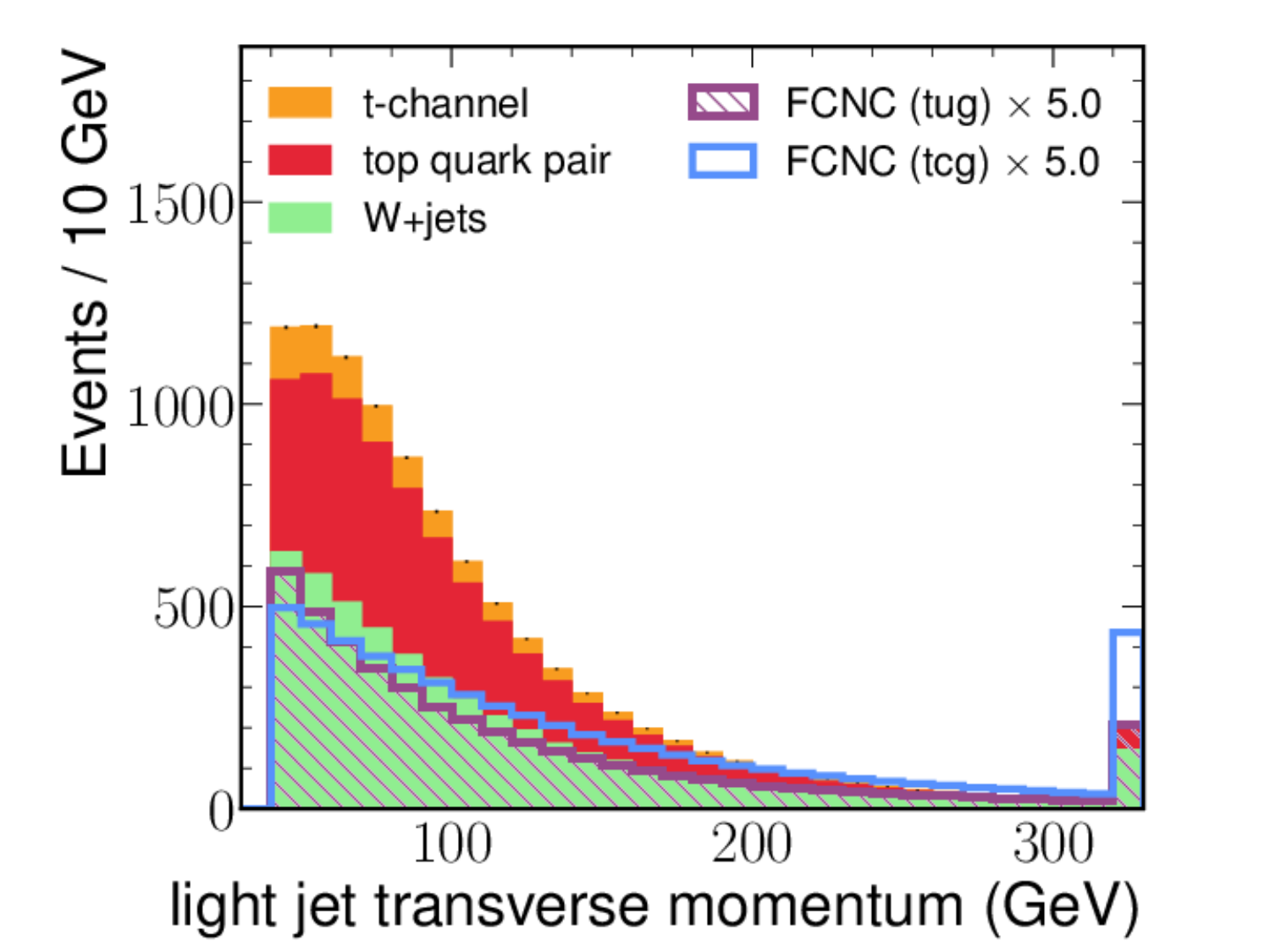}
  \includegraphics[width=\plotsizeHalf]{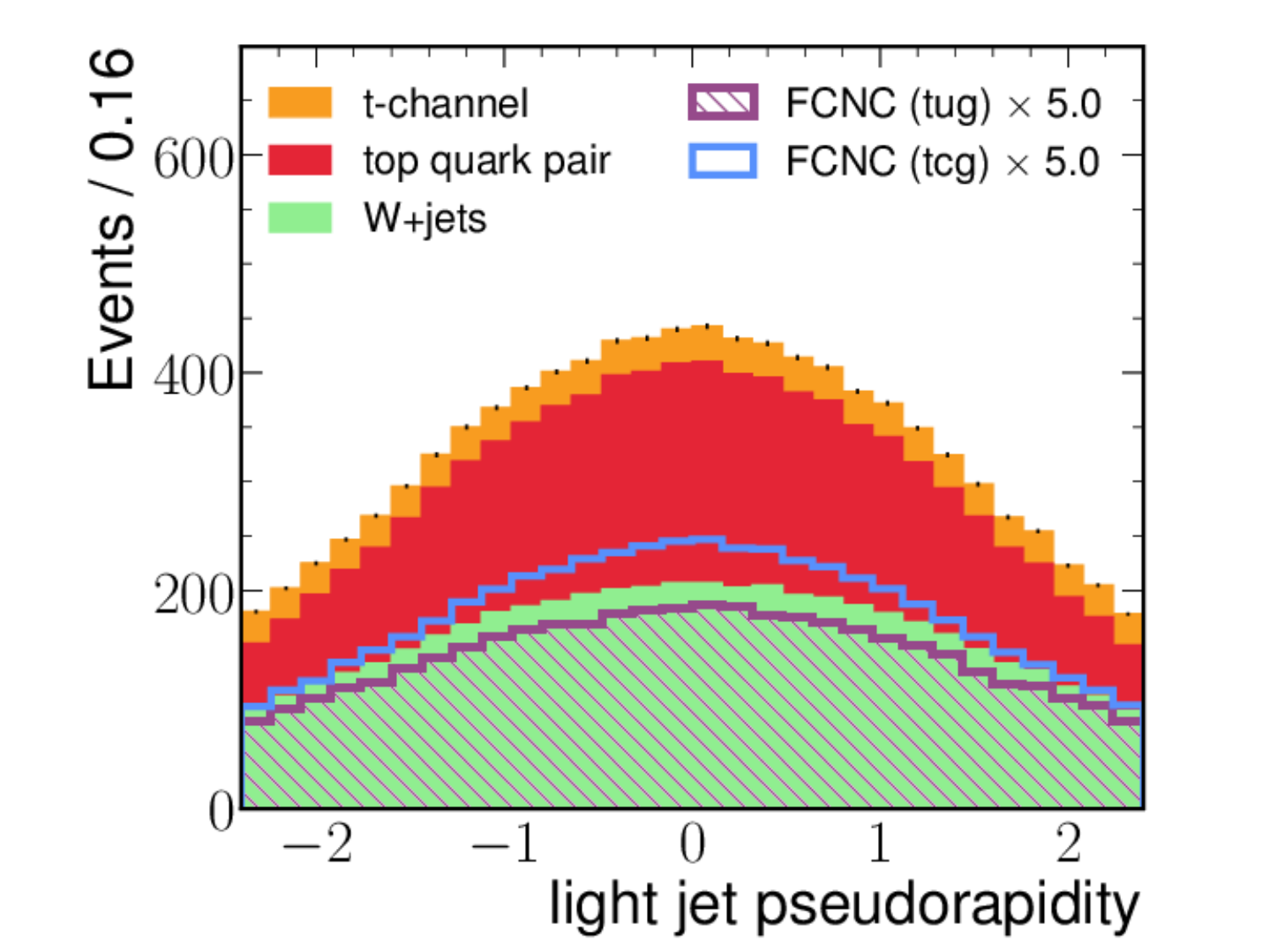}
  \caption{Distributions of reconstructed top quark kinematics and leading jet kinematics for the $tqg$ FCNC events and SM backgrounds. These variables are used as the input variables of the BDT and \SaJa. The total background statistical uncertainty is displayed by the black vertical lines. The distributions of $tug$ and $tcg$ FCNC events are drawn 5 times larger. The distributions are scaled to an integrated luminosity 138 fb$^{-1}$.}
  \label{fig:RecoTopAssoJetPlots}
\end{figure}

A previous search for $tqg$ FCNCs performed by the CMS Collaboration \cite{JHEP17CMStgq} used only the kinematic and event topology variables listed in \tbl{table:VarUsedBDT}.
The variables used are the kinematic variables of the reconstructed top quark and the leading jet and event topology variables such as opening angles in the $W$ boson and the top quark rest frames.
\fig{fig:RecoTopAssoJetPlots} displays kinematic distributions of the reconstructed top quark and the leading jet from $tqg$ FCNC events and SM backgrounds.

\begin{figure}
  \includegraphics[width=0.4\textwidth]{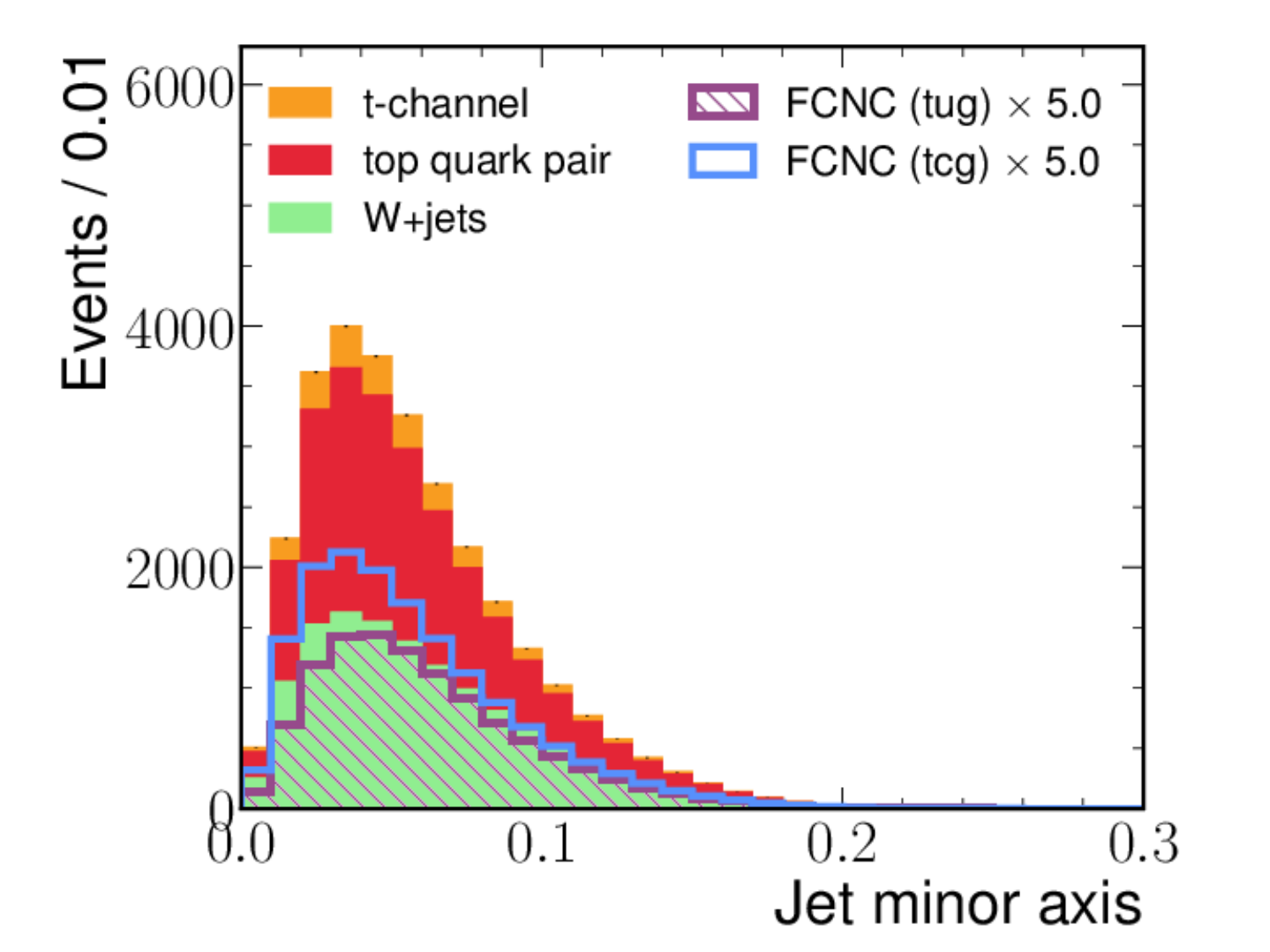}
  \includegraphics[width=0.4\textwidth]{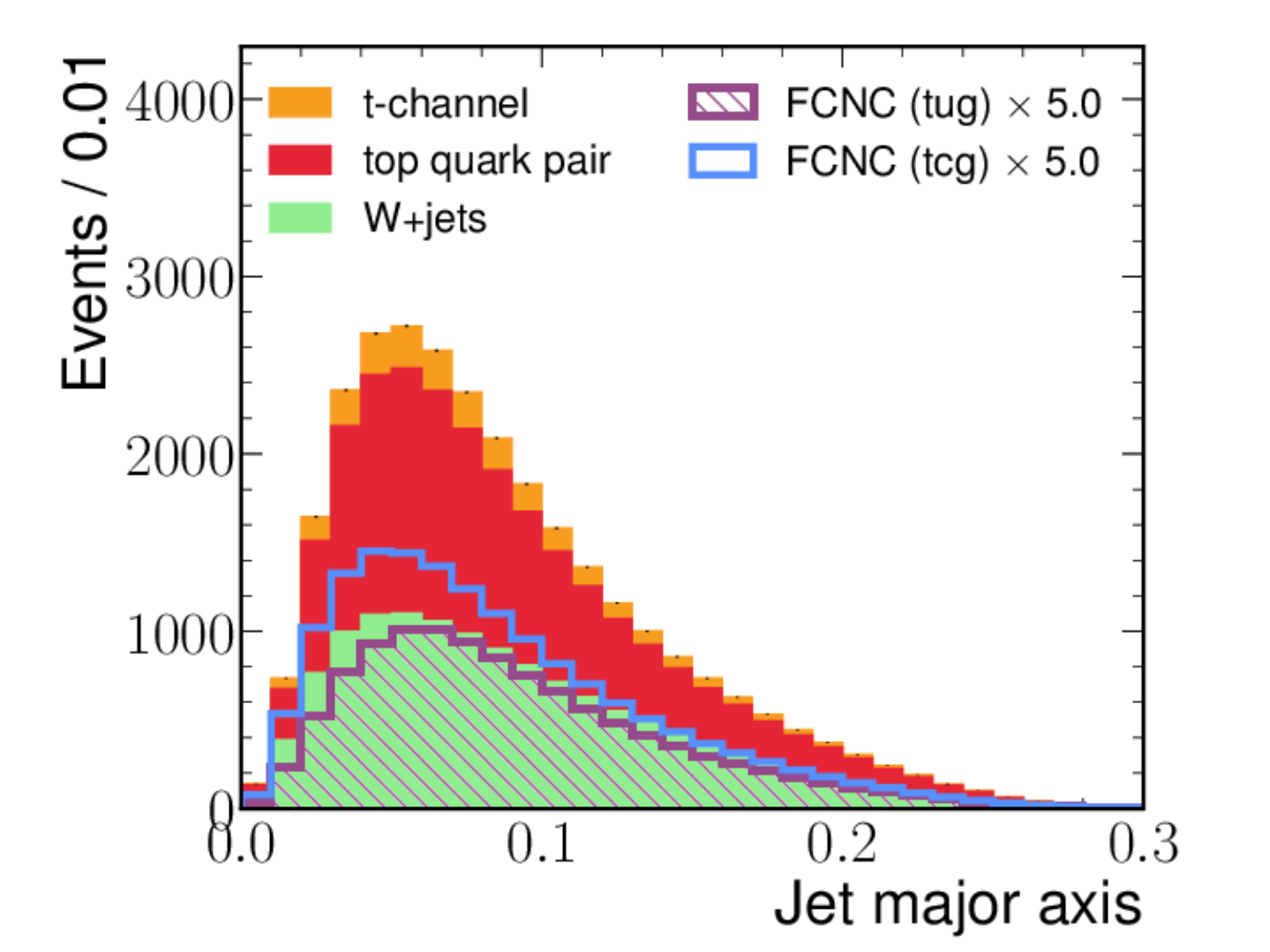}
                                                      
  \includegraphics[width=0.4\textwidth]{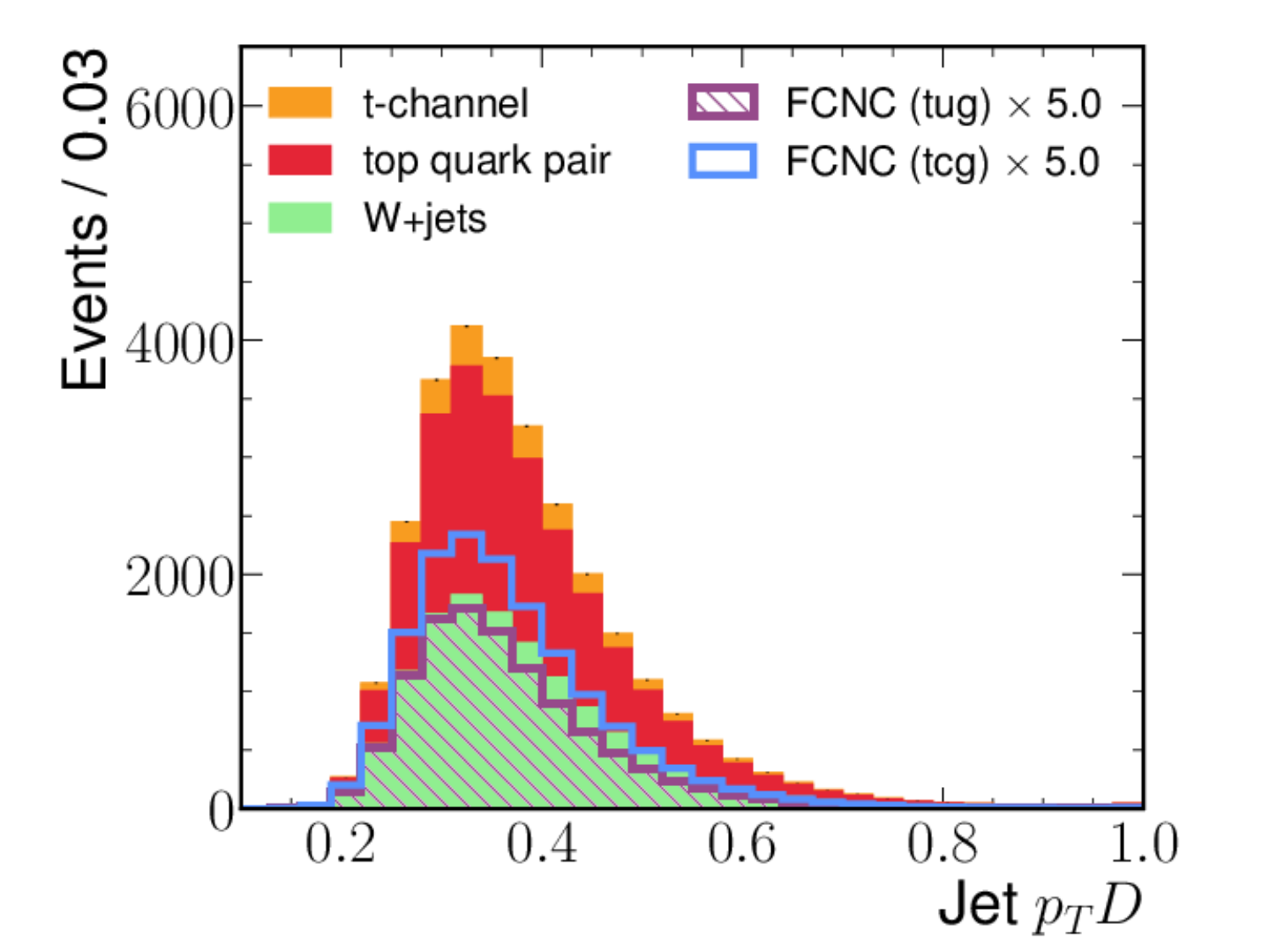}
  \includegraphics[width=0.4\textwidth]{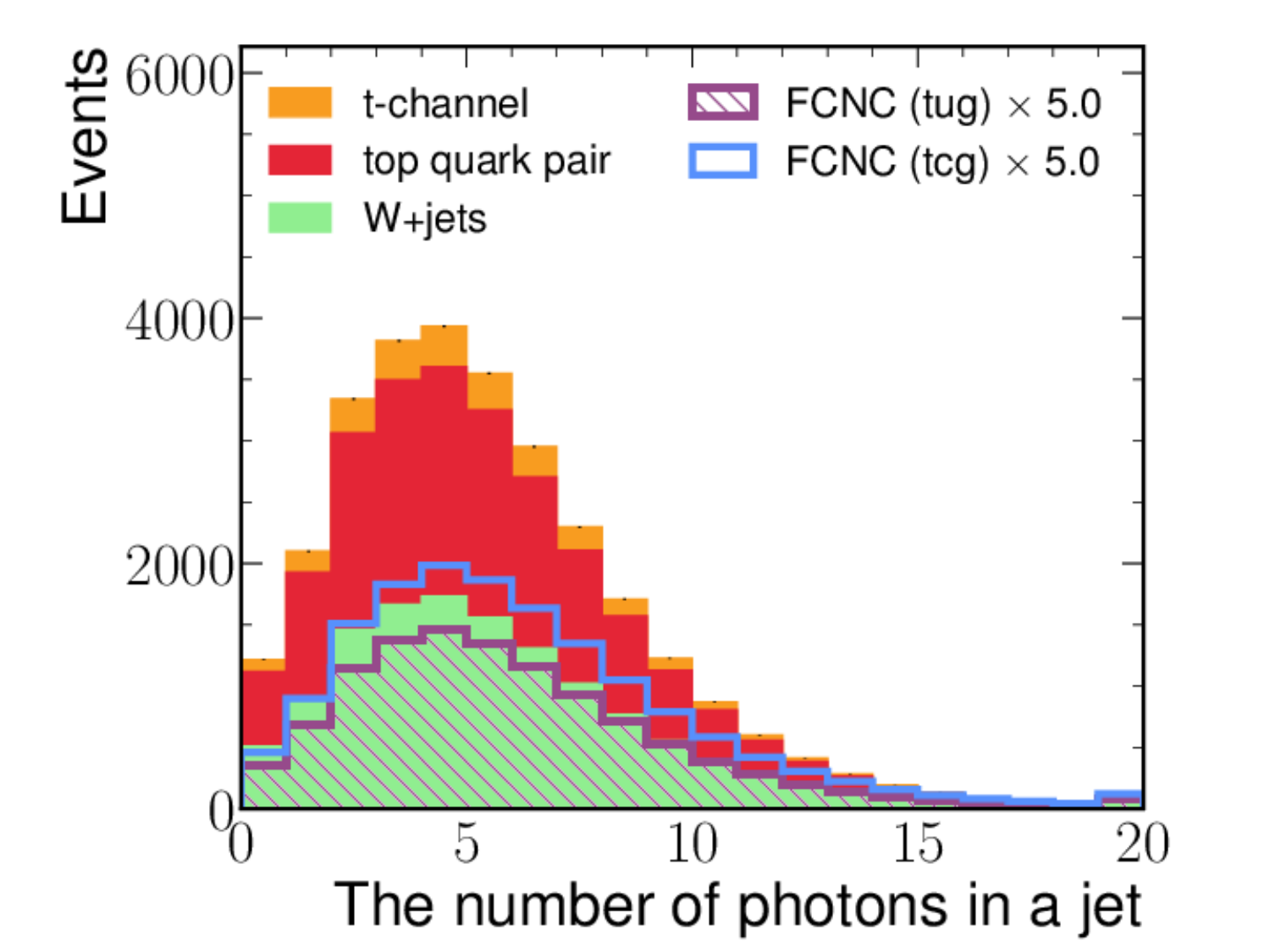}
                                                      
  \includegraphics[width=0.4\textwidth]{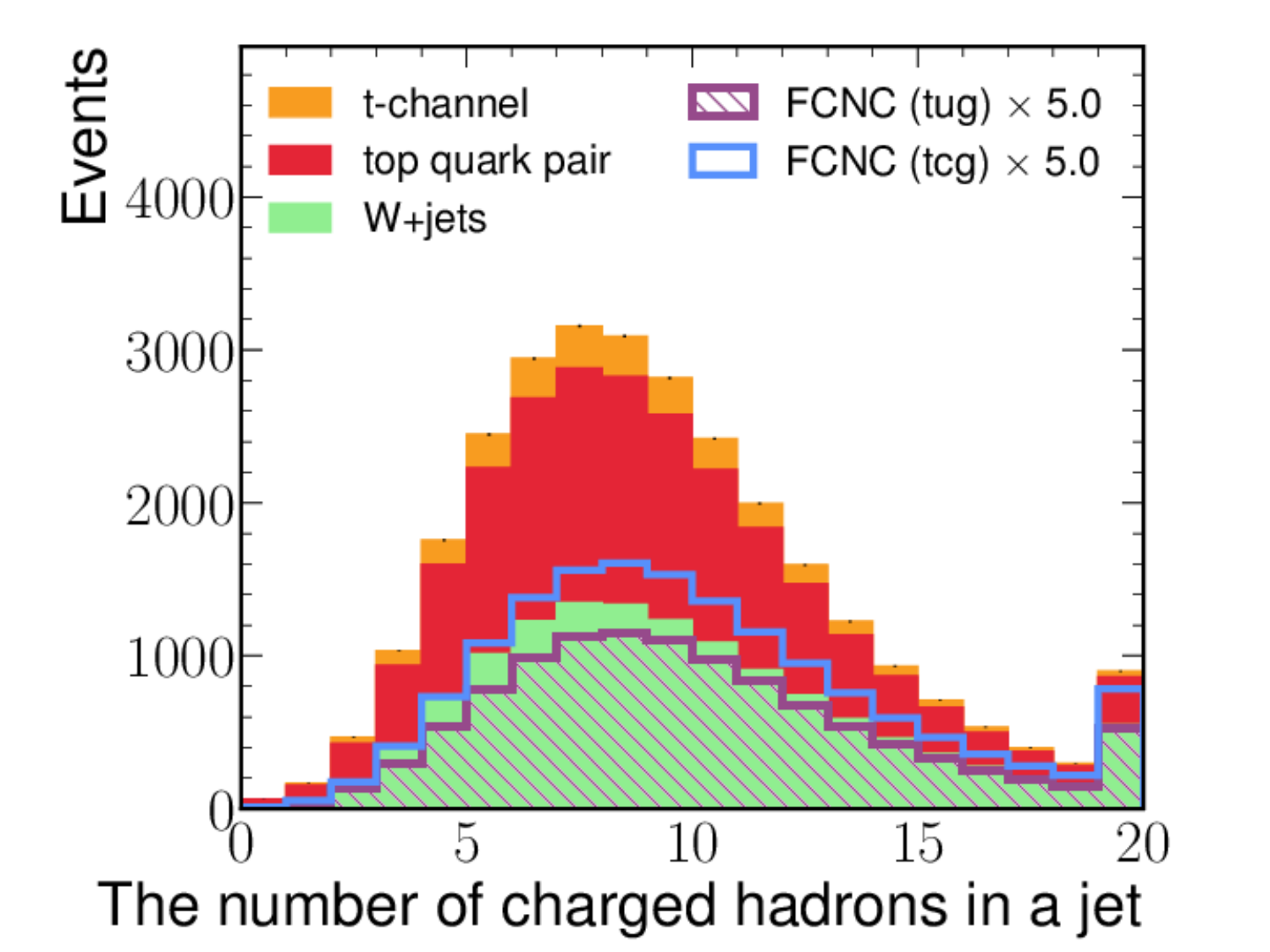}
  \includegraphics[width=0.4\textwidth]{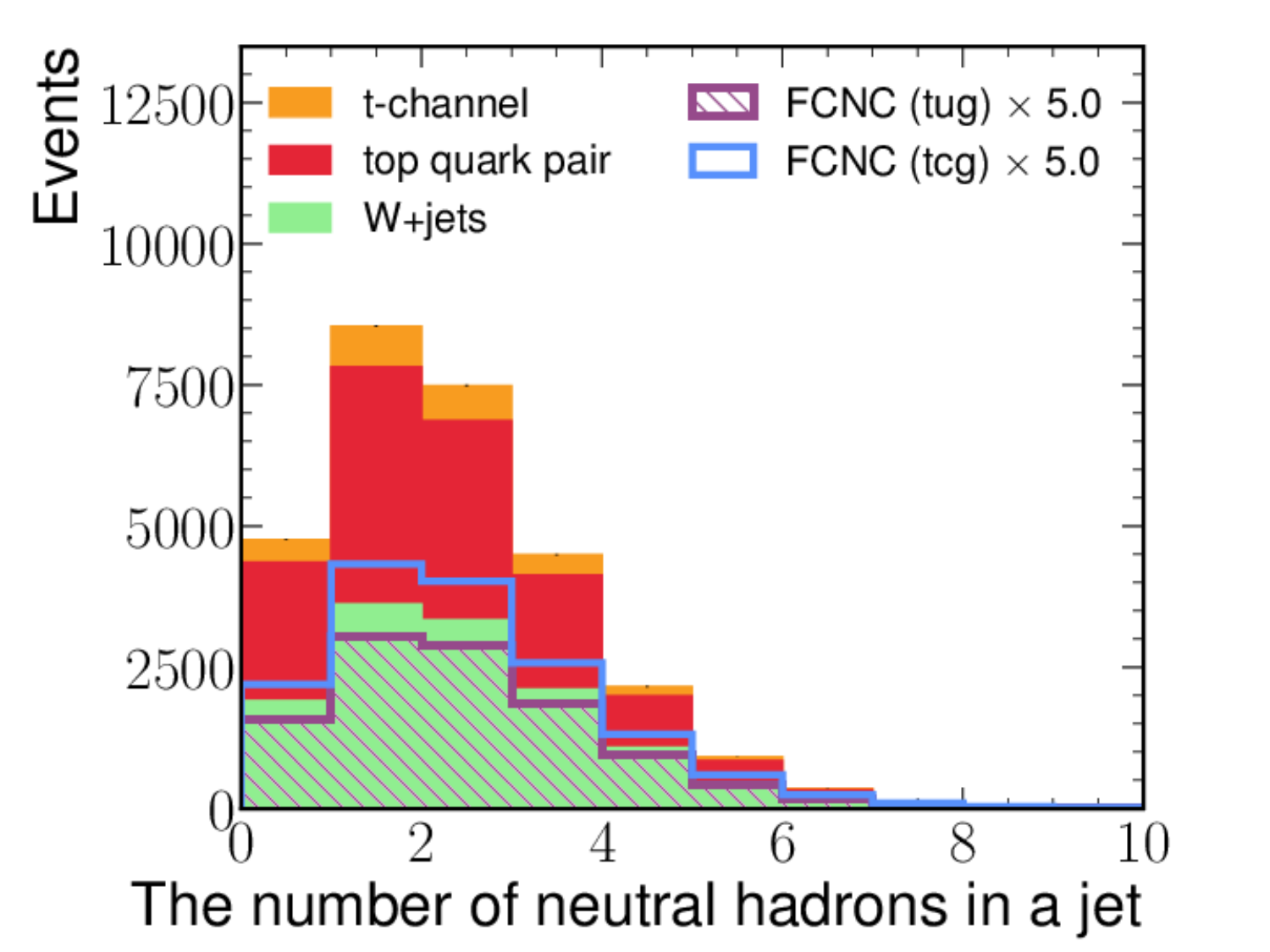}
                                                      
  \includegraphics[width=0.4\textwidth]{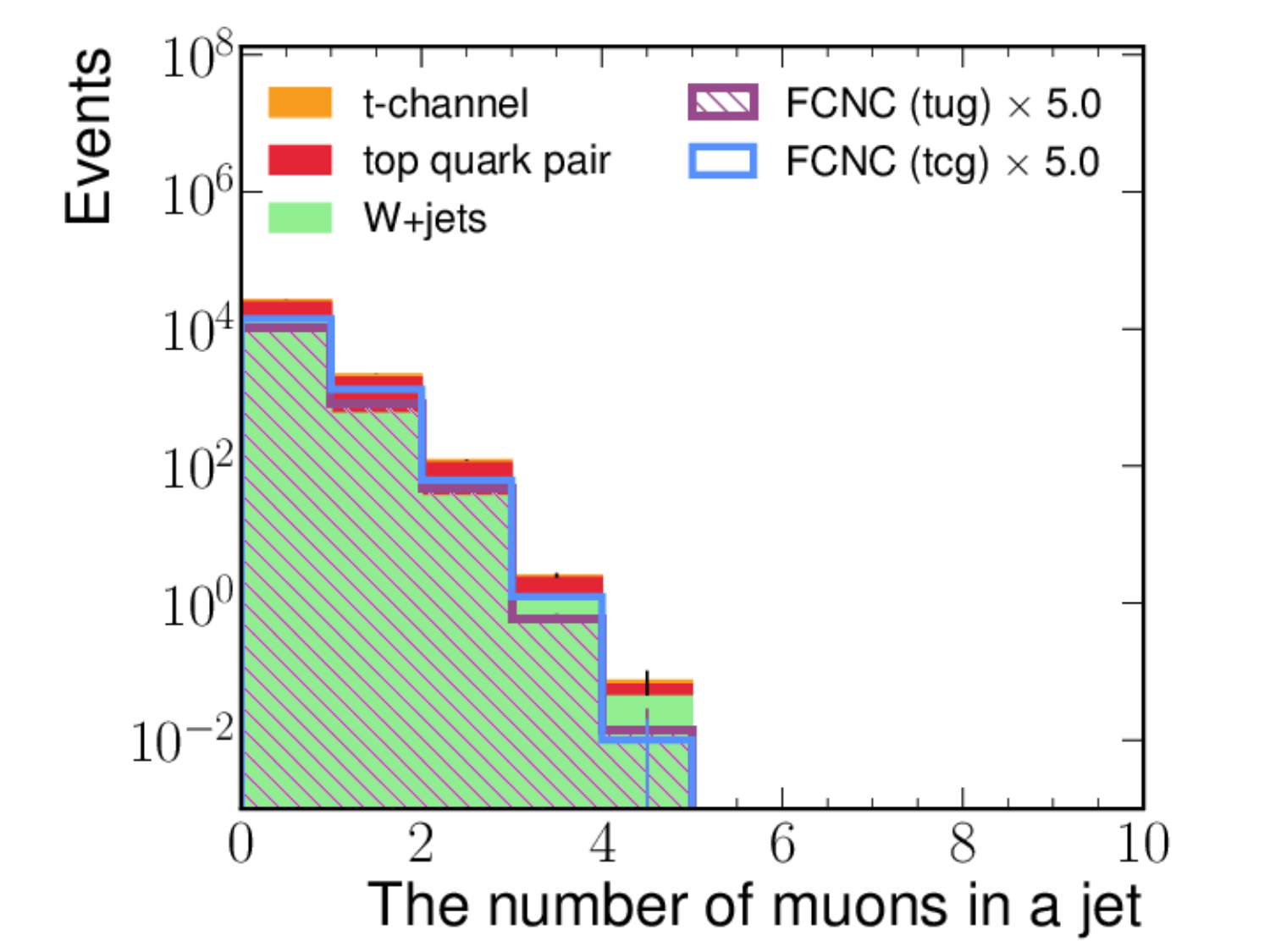}
  \includegraphics[width=0.4\textwidth]{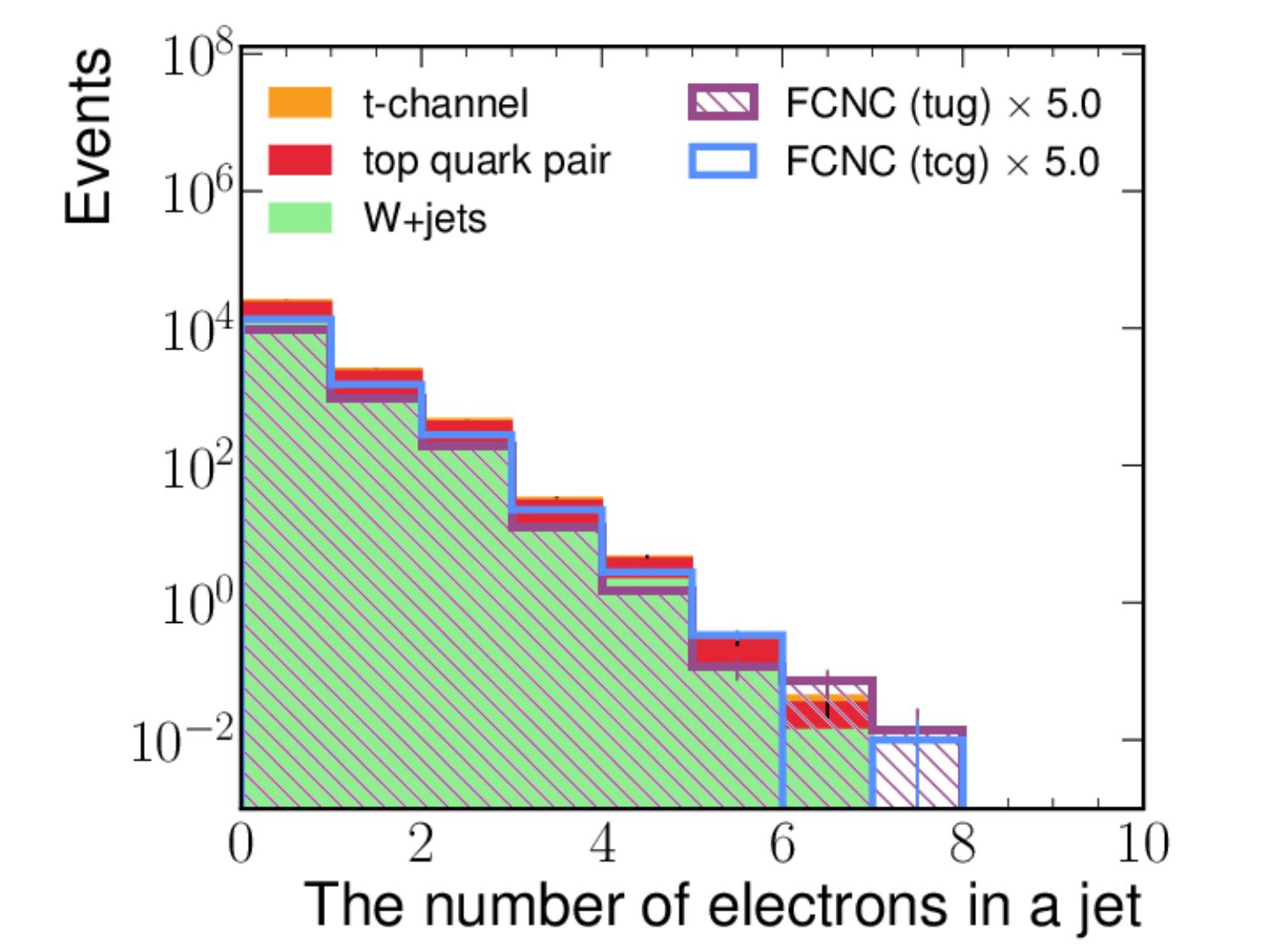}
  \caption{Distributions of the $qg$-discrimination variables for the $tqg$ FCNC events and SM backgrounds. These variables are used as the input variables of the BDT and \SaJa. The total background statistical uncertainty is displayed by the black vertical lines. The distributions of $tug$ and $tcg$ FCNC events are drawn 5 times larger. The distributions are scaled to an integrated luminosity 138 fb$^{-1}$.}
  \label{fig:QGDiscPlots}
\end{figure}

The higher color factor of gluons compared to quarks results in gluon jets generally having more jet constituents, broader shape, and softer fragmentation compared to quark jets.
The CMS Collaboration has performed studies of $qg$-discrimination using the multiplicity of jet constituents, major (minor) axes $\sigma_M(j)$ ($\sigma_m(j)$) of jets in the $\eta-\phi$ space, and jet energy sharing $p_TD = \frac{\sqrt{\sum{p_{T,i}^2}}}{\sum{p_{T,i}}}$, where $i$ indexes over jet constituents, as inputs \cite{CMSQGDiscPAS,CMSQGDiscProceeding}. 
Further studies suggest splitting the total multiplicity to the individual numbers of charged hadrons, neutral hadrons, electrons, muons, and photons \cite{QGDiscSYCNN,QGDiscSYWSL}.
The $gq$-discrimination variables used in this paper are listed in \tbl{table:VarQGDisc}, and \fig{fig:QGDiscPlots} shows their distributions.

\begin{table}[b]
  \begin{tabular}{ll}
    Variables & Definition \\
    \hline
    \hline
    $\eta(l)$       & Pseudorapidity of the lepton \\
    $p_T(j_L)$      & Transverse momentum of the leading light jet \\
    $\eta(j_L)$     & Pseudorapidity of the leading light jet \\
    $p_T(b)$        & Transverse momentum of the leading $b$-tagged jet \\
    $p_T(j_1)$      & Transverse momentum of the leading jet \\
    sgn$(l)$        & Charge of the lepton \\
    $H_T(j_1, j_2)$ & The scalar sum of $p_T$ of the leading jet and the sub-leading jet \\
    $p_T(j_1, j_2)$ & The vector sum of $p_T$ of the leading jet and the sub-leading jet \\
    $m(j_1, j_2)$   & The invariant mass of the leading jet and the sub-leading jet  \\
    $m(W+\sum j)$   & The invariant mass of the $W$ boson and jets \\
    $p_T(t)$        & Transverse momentum of the top quark \\
    $m(t)$          & The mass of top quark \\
    Planarity       & \shortstack{The smallest eigenvalue of a tensor $\left( \sum_\alpha p_\alpha^i p_\alpha^j \right) / \left( \sum_\alpha (p_\alpha)^2 \right)$, \\ where the summations run over jets, the lepton and \vecmet} \\
    $\cos{\theta(l, j)_{\textrm{top}}}$   & \shortstack{Opening angle of the lepton and the leading light jet \\ in the top quark rest frame} \\
    $\cos{\theta(l, W)_{\textrm{W}}}$     & \shortstack{Opening angle of the lepton and the $W$ boson direction \\ in the $W$ boson rest frame} \\
    $\cos{\theta(W, j)_{\textrm{top}}}$   & \shortstack{Opening angle of the $W$ boson and the leading light jet \\ in the top quark rest frame} \\
  \end{tabular}
  \caption{The input variables of the BDT. This includes information from the lepton, the missing transverse momentum, the leading jet, the leading $b$-tagged jet, the $W$ boson, and the top quark. 
  }
  \label{table:VarUsedBDT}
\end{table}

\begin{table}
  \begin{tabular}{ll}
    Variables & Definition \\
    \hline
    $N_{h\pm}(j)$                          & The number of charged hadrons in jet $j$ \\
    $N_{h0}(j)$                            & The number of neutral hadrons in jet $j$ \\
    $N_{e}(j)$                             & The number of electrons in jet $j$ \\
    $N_{\mu}(j)$                           & The number of muons in jet $j$ \\
    $N_{\gamma}(j)$                        & The number of photons in jet $j$ \\
    $\sigma_{M}(j)$, $\sigma_{m}(j)$       & The major and minor axes of jet $j$ in the $\eta-\phi$ space \\
    $p_T D(j)$                             & $\sqrt{\sum_i (p_{T,i})^2} / \sum_i p_{T,i}$ ($i$ is indexed over the jet constituents) \\
  \end{tabular}
  \caption{The $qg$-discrimination variables that are input into the machine learning models. The variables are taken from \cite{QGDiscSYWSL}.}
  \label{table:VarQGDisc}
\end{table}

The impact of the $qg$-discrimination variables is evaluated by comparing two BDTs.
The first BDT is constructed with variables listed in \tbl{table:VarUsedBDT}, and the other one is constructed with variables in both \tbl{table:VarUsedBDT} and \ref{table:VarQGDisc}.
We train BDTs to discriminate $tqg$ FCNC signal from SM backgrounds.
We use Toolkit for Multivariate Data Analysis (TMVA) v.4.3.0, which is a part of the ROOT framework, to implement the BDTs with AdaBoost \cite{TMVA, AdaBoost, ROOT}.
The parameters used for the BDTs are listed in Table~\ref{table:BDThyperparameters}.
During BDT training, we weight the background events such that each background sample is normalized to its cross-section.
Additionally, we scale the total background weights to match the number of signal events.
We have four BDTs.
The $tug$ and $tcg$ signal samples are separately trained with a model that uses the input variables listed in only \tbl{table:VarUsedBDT} as the baseline, and another model that includes the baseline variables and $qg$-discrimination variables in \tbl{table:VarQGDisc} for the leading jet.

\begin{table}[h]
  \begin{tabular}{lll}
    Name & \;\; Description & \;\; Value \\
    \hline
    \hline
    \texttt{AdaBoostBeta}        & \;\; Learning rate for AdaBoost algorithm                                                   & \;\; 0.5 \\
    \texttt{nCuts}                & \;\; \shortstack{Number of grid points in variable range used in \\finding optimal cut in node splitting} & \;\; 20 \\
    \texttt{NTrees}               & \;\; Number of trees in the forest                                                          & \;\; 850 \\
    \texttt{MinNodeSize}          & \;\; Minimum percentage of training events required in a leaf node                          & \;\; 2.5\% \\
    \texttt{MaxDepth}             & \;\; Max depth of the decision tree allowed                                                 & \;\; 3 \\
    \texttt{BaggedSampleFraction} & \;\; \shortstack{Relative size of bagged event sample to \\original size of the data sample}              & \;\; 0.5 \\
    \hline
  \end{tabular}
  \caption{The hyperparameters of all BDTs. The variable names and the descriptions are from the TMVA manual \cite{TMVA}}
  \label{table:BDThyperparameters}
\end{table}

In the BDT setup, we assume that the leading jet (i.e., the most energetic jet which is not $b$-tagged) comes from the associated parton.
After the event selection, the associated parton is the leading jet in 60\% of events, using $\Delta R < 0.3$ to match jets to partons.
Unmatched partons are primarily due to initial state radiation, mistagging the $b$-jet from the top quark decay, or gluon splitting.
As the BDT model takes in a fixed number of inputs, and cannot adapt to varying numbers of jets, we use only the leading jet variables in this study.
This limits the performance of the BDT due to the misidentification of the associated parton.
As an alternative approach which solves this issue, we adopt the deep learning model \SaJa\, which can take an arbitrary number of jets as input.

\begin{figure}[h]
  \includegraphics[width=15.0cm]{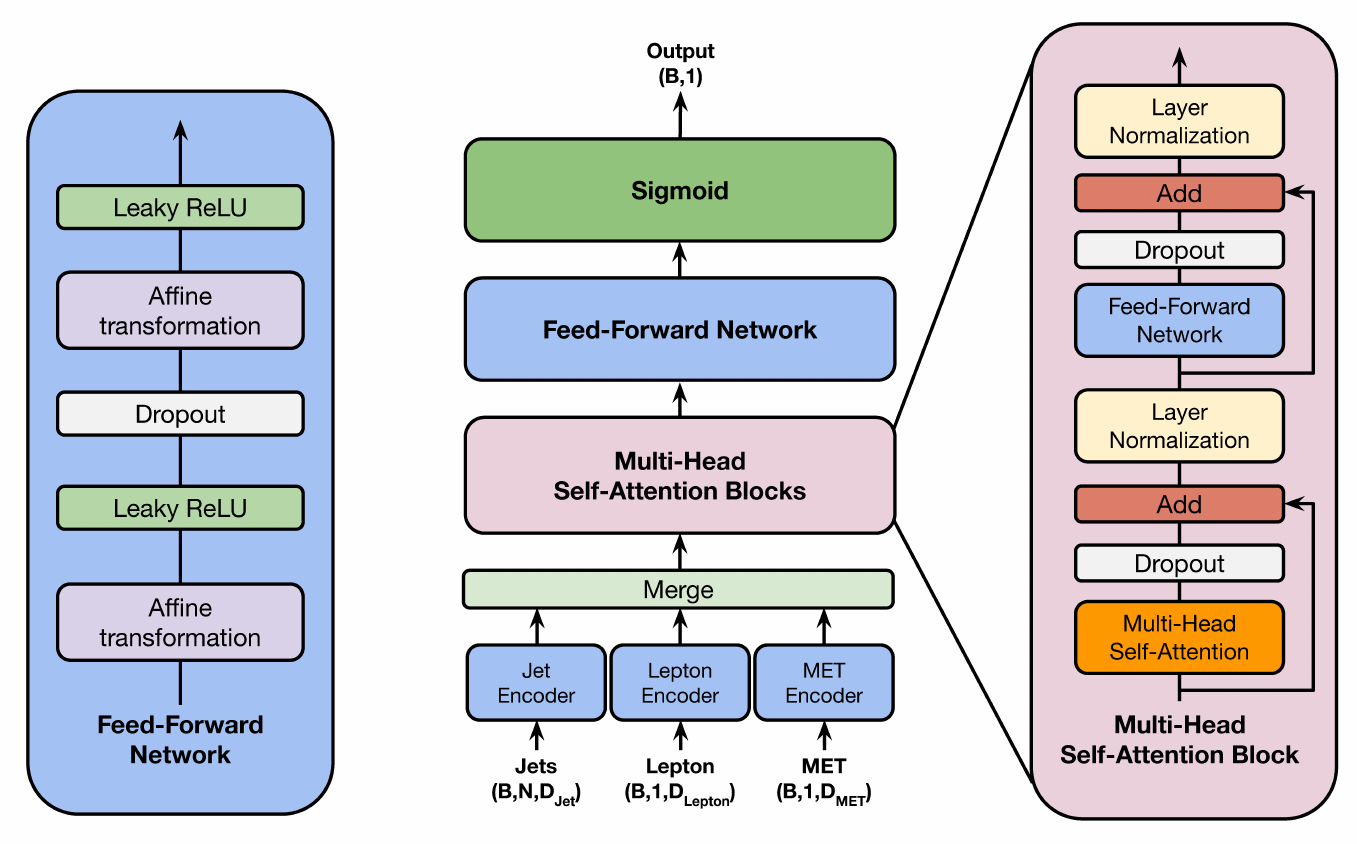}
  \caption{Diagram showing the data flow of the \SaJa\ model used in this study, where $B$ is the batch size, $N$ is the number of jets and $D_{\textrm{Jet}}$, $D_{\textrm{Lepton}}$, $D_{\textrm{MET}}$ are the number of input variables of each jet, the lepton, and the missing transverse momentum, respectively. These number are determined by the number of input variables in \tbl{table:VarQGDisc} and \ref{table:VarUsedSaJa}.}
  \label{fig:SaJaModel}
\end{figure}

\SaJa\ is a Transformer-based model \cite{AttentionIsAllYouNeed} which consists mainly of feed-forward networks and multi-head self-attention blocks.
The original \SaJa\ model is designed to match jets to partons, so it only takes jets as inputs. 
In this analysis, however, we extend the input objects to include the lepton and the missing transverse momentum as well as jets.
For this purpose, we allocate an encoder, defined as a feed-forward network, to each input physics object, which takes the object's input variables to a vector of a common size.
The dimensions of the input vectors corresponding to each object are $D_{\textrm{Lepton}} = 6$, $D_{\textrm{MET}} = 2$, and $D_{\textrm{Jet}} = 13$ when the $qg$-discrimination variables are used, otherwise $D_{\textrm{Jet}} = 5$ (per \tbl{table:VarQGDisc} and \ref{table:VarUsedSaJa}).
The output of the encoders is concatenated and fed into a sequence of multi-head self-attention blocks.
All operations have the same output dimension, denoted as $D_{\text{model}}$, except the output dimension of the first affine transformation in all feed-forward networks, denoted as $D_{\text{feedforward}}$.
The structure of \SaJa\ in this study is depicted in \fig{fig:SaJaModel}.

The output of the original \SaJa\ model is jet-wise scores of assignment to partons, so the original model has jet-wise feed-forward networks and softmax after the self-attention blocks.
Since our model is designed to classify $tqg$ FCNC events from SM backgrounds, we construct a single event latent vector by taking the mean along the object axis and take one feed-forward network followed by a sigmoid $\frac{1}{1 + e^{-x}}$ giving a classification score between 0 and 1 as the model output.

We consider five hyperparameters for \SaJa, $D_{\text{feedforward}}$, $D_{\text{model}}$, the number of heads, the number of blocks in the sequence of self-attention blocks, and the dropout \cite{Dropout} rate.
The specific values for these hyperparameters are listed in \tbl{table:SaJaHyperparameters}.
We use Adam optimization algorithm \cite{Adam}, where the learning rate is set to 0.0003, $\beta_1 = 0.9$, and $\beta_2 = 0.999$.
For both BDT and \SaJa, we separate $tqg$ FCNC events and SM backgrounds into the training dataset (64\%), test dataset (20\%), and validation dataset (16\%).
When training each model, we use the same training, test, and validation datasets.
During training, we weight the background events as in the BDT case.
As for the case of the BDT training, $tug$ and $tcg$ are trained separately with a baseline input variable model (listed in \tbl{table:VarUsedSaJa} for \SaJa) and a model including $qg$-discrimination variables (\tbl{table:VarQGDisc}).
Since \SaJa can combine the information from the individual input objects through the self-attention layers to produce a representation which can be efficiently used for event classification, the input variables for \SaJa\ are only the physics object-related variables, whereas the BDT variables include composite variables, like the reconstructed top quark mass and planarity.

\begin{table}[h]
  \begin{tabular}{ll}
    Name & \;\; Value \\
    \hline
    \hline
    $D_{\text{feedforward}}$            & \;\; 256 \\
    $D_{\text{model}}$                  & \;\; 160 \\
    The number of heads                 & \;\; 10 \\
    The number of self-attention blocks & \;\; 2 \\
    Dropout rate                        & \;\; 0.1 \\
    \hline
  \end{tabular}
  \caption{The hyperparameters of all \SaJa\ networks}
  \label{table:SaJaHyperparameters}
\end{table}

\begin{table}[h]
  \begin{tabular}{ll}
    Variables & Definition \\
    \hline
    \hline
    $p_T(j)$, $\eta(j)$, $\phi(j)$, $m(j)$ & Kinematic variables of each jet \\
    $f_{b}$                                & A variable with value 1 (0) if the jet is (is not) $b$-tagged \\
    \hline
    $p_T(l)$, $\eta(l)$, $\phi(l)$, $m(l)$ & Kinematic variables of the lepton \\
    sgn$(l)$ & Charge of the lepton \\
    muon bit & A variable with value 1 (0) if the lepton is muon (electron) \\
    \hline
    $\MET$, $\phi(\MpT)$ & Magnitude and azimuthal angle of \vecmet \\
  \end{tabular}
  \caption{The input variables of the \SaJa\ network. This includes the information on the lepton, missing transverse momentum, and all jets in the event.
  }
  \label{table:VarUsedSaJa}
\end{table}

\section{Results}
\label{sec:result}
\begin{figure}[ph]
  \includegraphics[width=\plotsizeHalf]{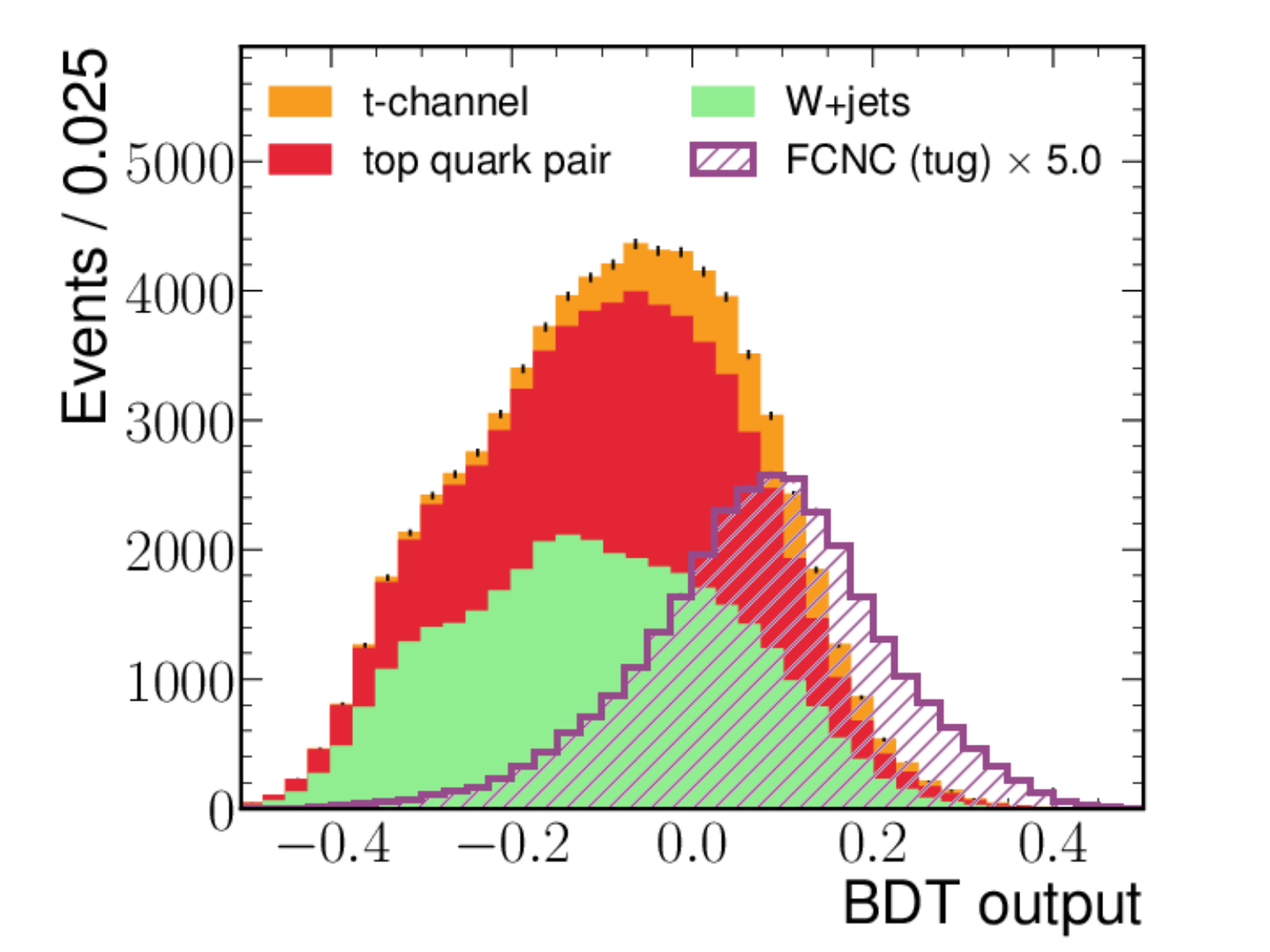}
  \includegraphics[width=\plotsizeHalf]{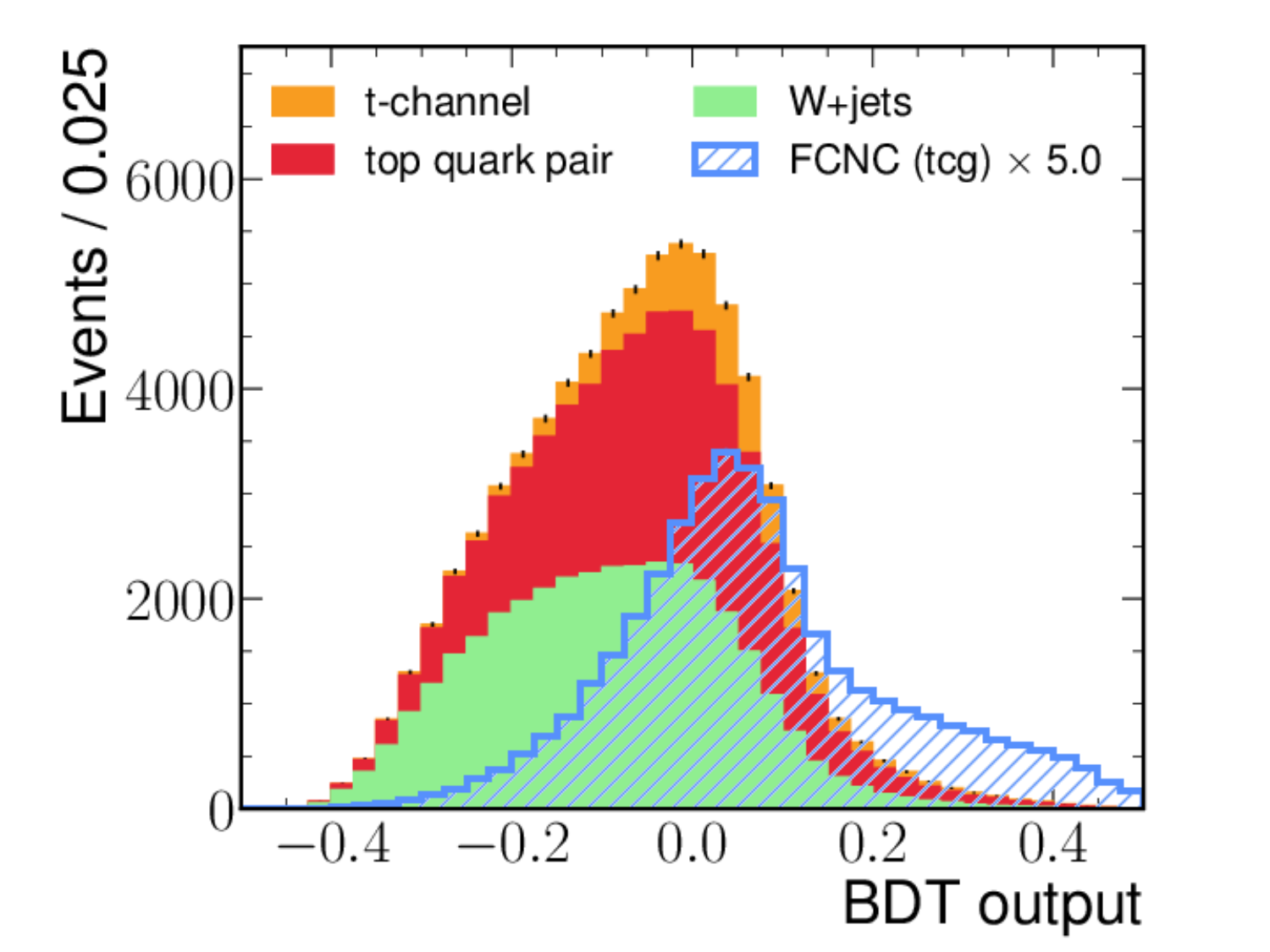}

  \includegraphics[width=\plotsizeHalf]{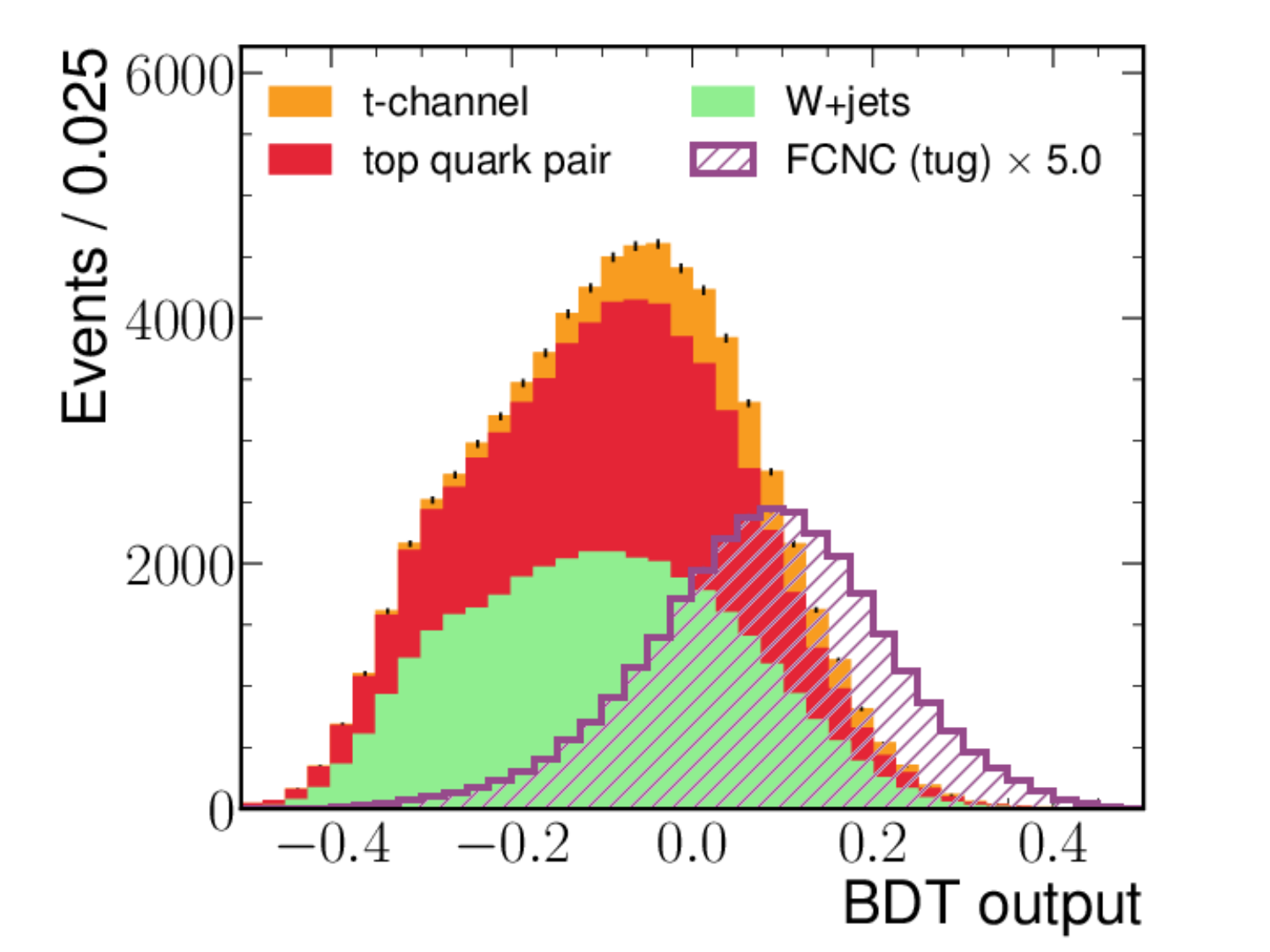}
  \includegraphics[width=\plotsizeHalf]{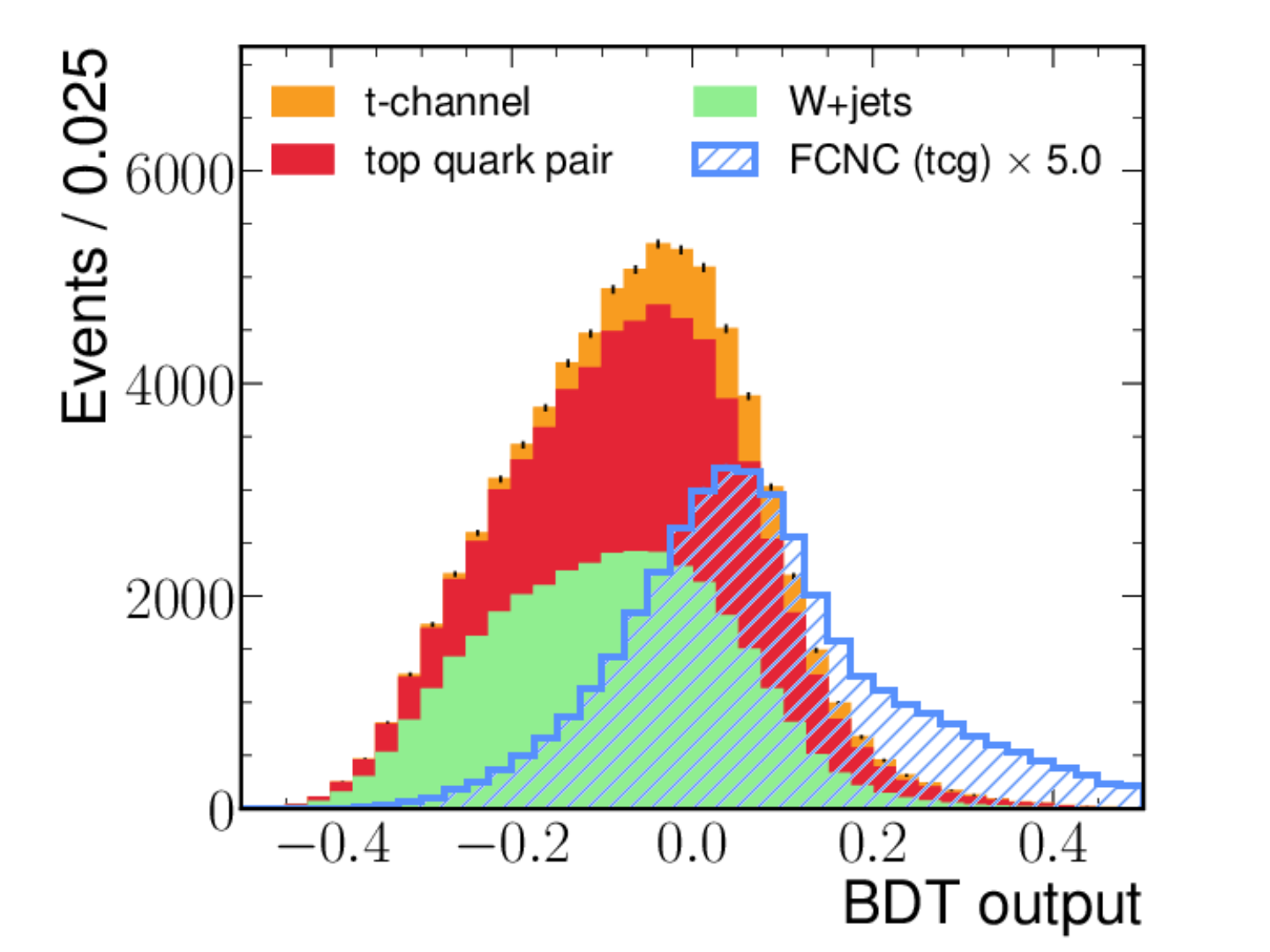}
  \caption{
    The output distribution of BDTs obtained from the test dataset.
    The left (right) columns are from the training with $tug$ ($tcg$) signal samples.
    The networks for the top row use the baseline input variables, and those for the bottom row add the $qg$-discrimination variables.
    The total background statistical uncertainty is displayed by the black vertical lines.
  }
\label{fig:ScoreDistBDT}
\end{figure}

\begin{figure}[ph]
  \includegraphics[width=\plotsizeHalf]{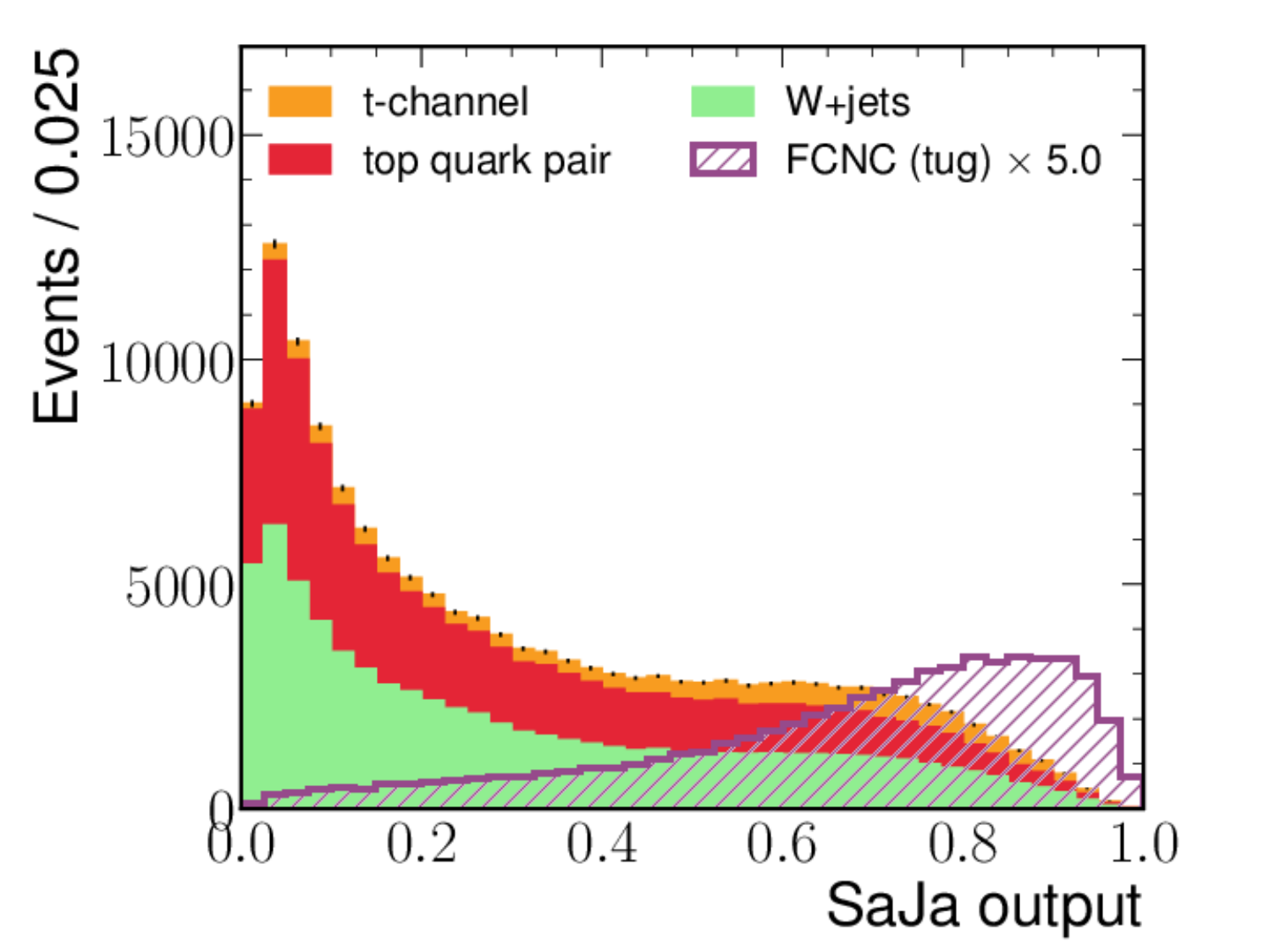}
  \includegraphics[width=\plotsizeHalf]{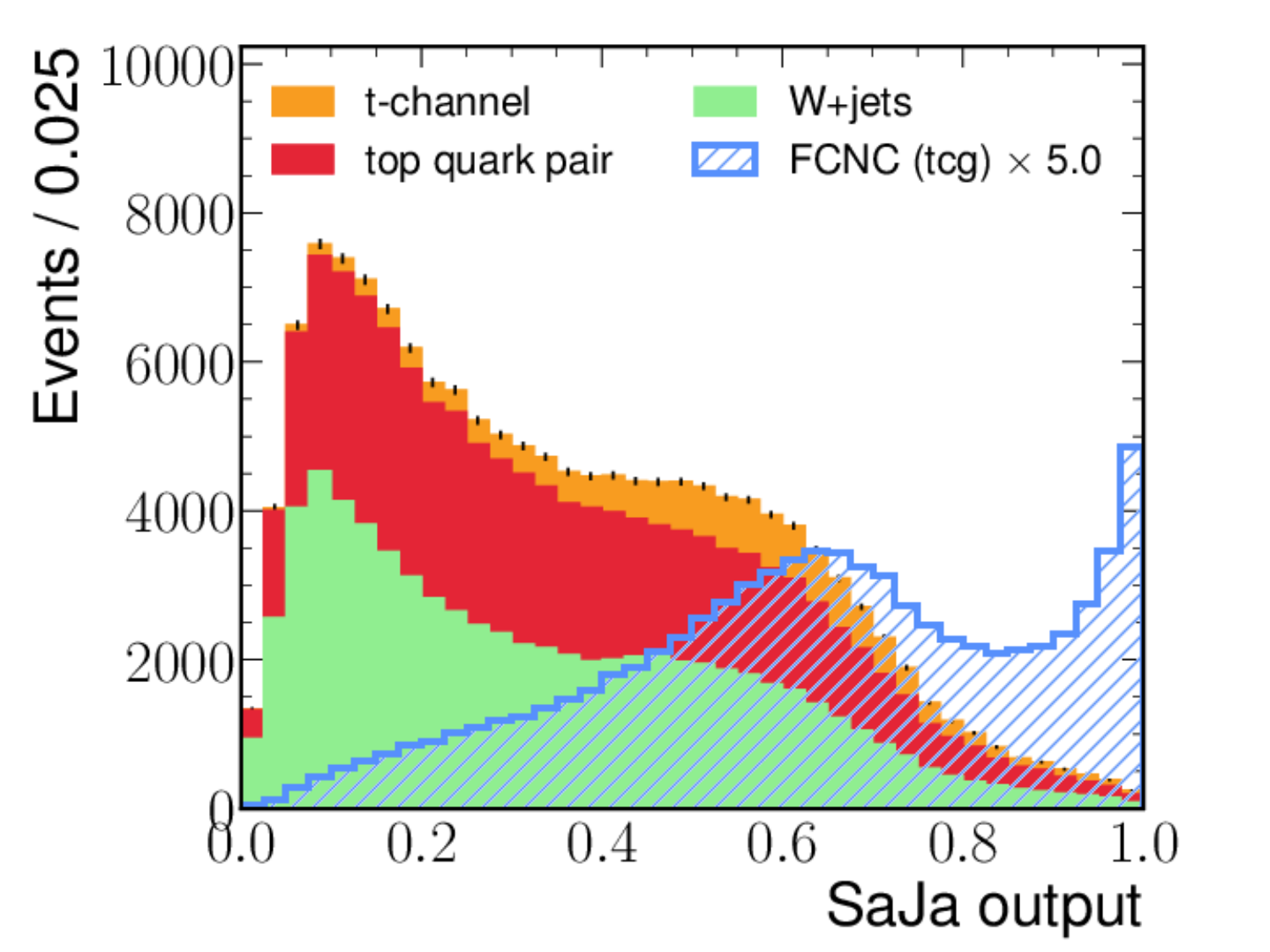}
  
  \includegraphics[width=\plotsizeHalf]{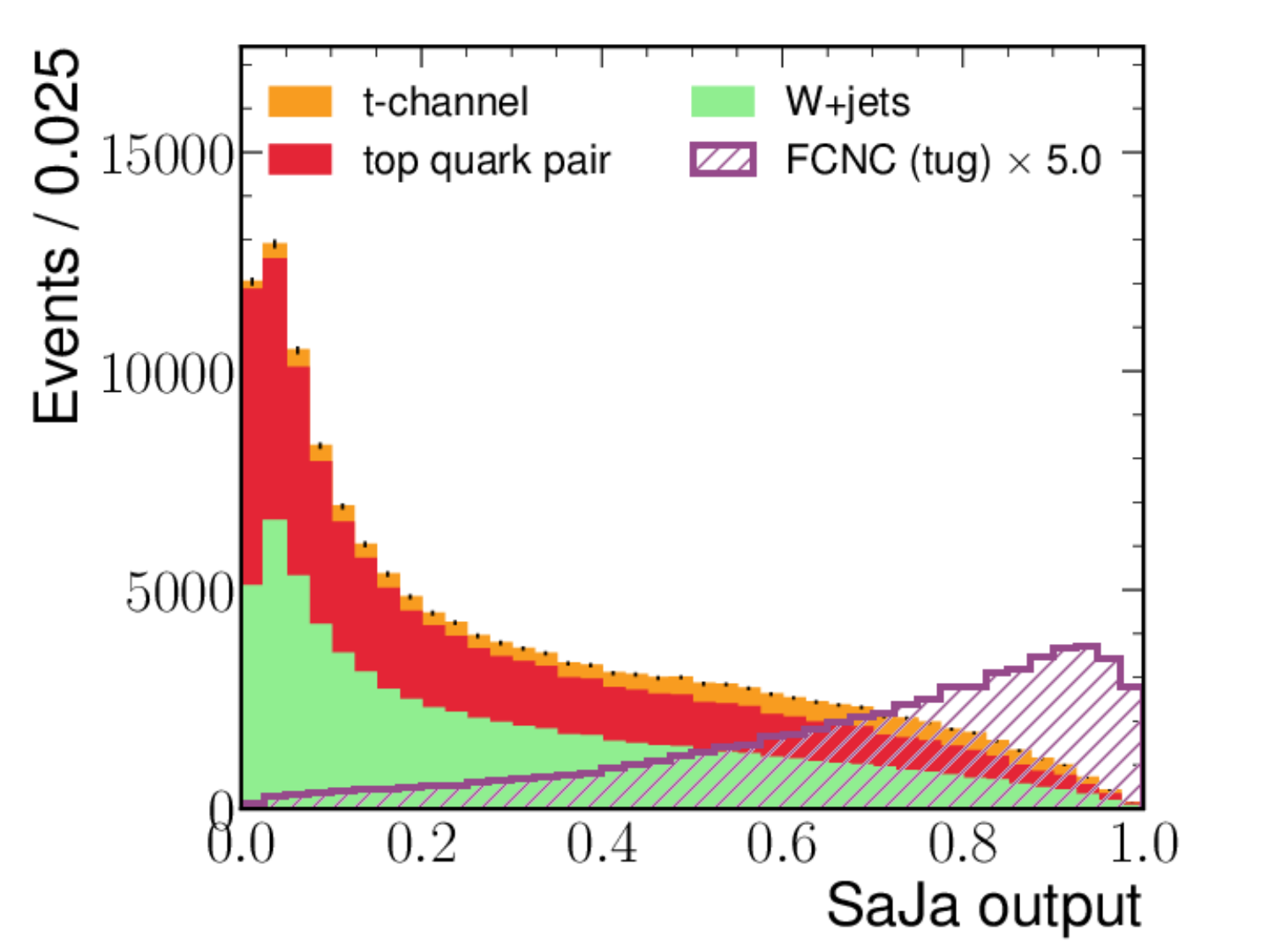}
  \includegraphics[width=\plotsizeHalf]{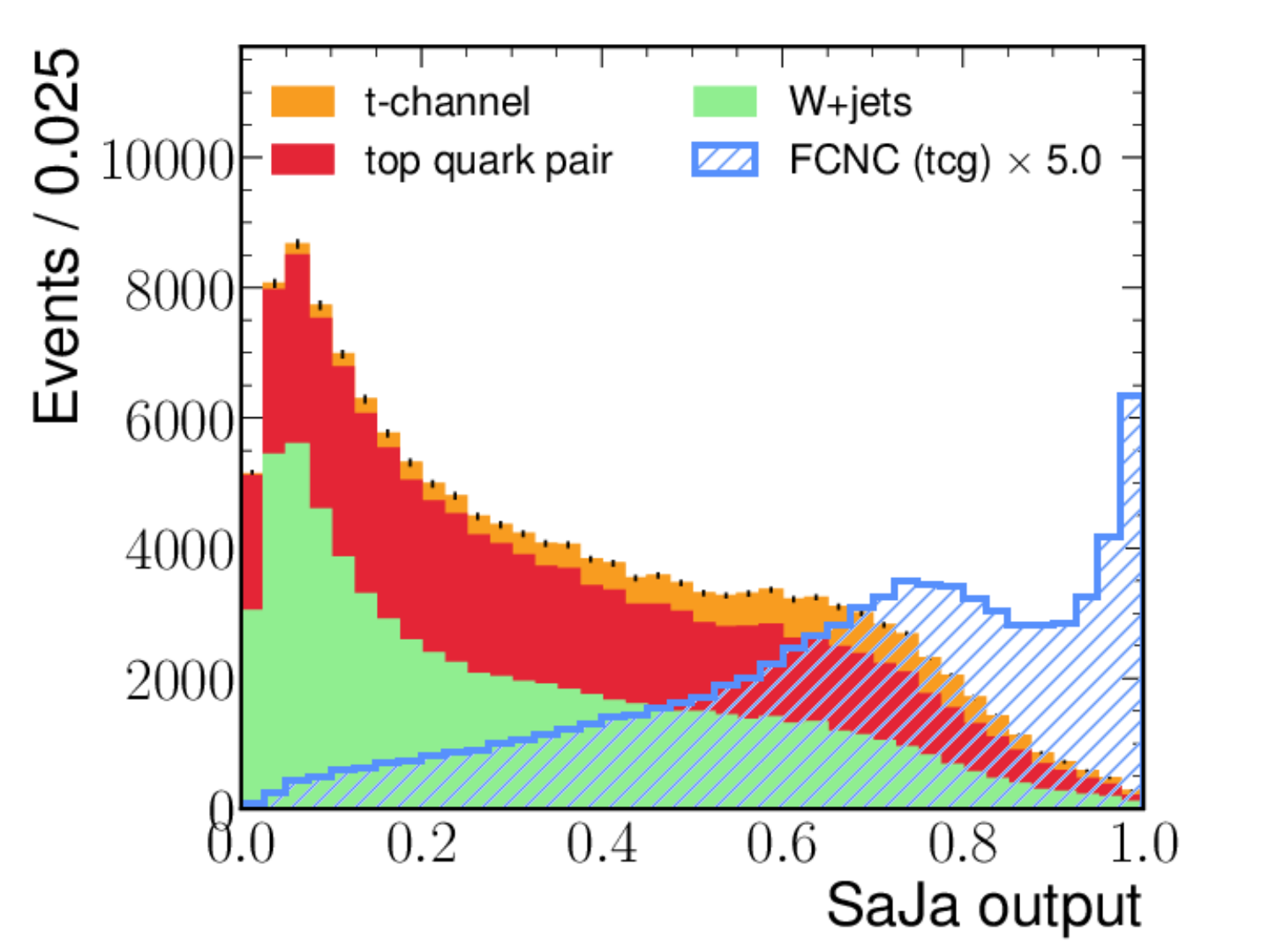}
  \caption{
    The output distribution of \SaJa\ networks obtained from the test dataset.
    The left (right) columns are from the training with $tug$ ($tcg$) signal samples.
    The networks for the top row use the baseline input variables, and those for the bottom row add the $qg$-discrimination variables.
    The total background statistical uncertainty is displayed by the black vertical lines.
  }
\label{fig:ScoreDistSaJa}
\end{figure}

The output distributions of the BDTs and \SaJa\ networks, obtained from the test dataset, are displayed in \fig{fig:ScoreDistBDT} and \fig{fig:ScoreDistSaJa}.
For each of the output distributions, we construct the receiver operating characteristic (ROC) curve and the significance improvement characteristic (SIC) \cite{JHEP11BlackSIC} as a function of the signal efficiency.
The SIC is defined as $\epsilon_S / \sqrt{\epsilon_B}$, where $\epsilon_S$ is the signal efficiency and $\epsilon_B$ is the background efficiency.
We compare the methods by integrating the ROC curve to get the area under the curve (AUC) and by finding the maximum SIC.

\begin{table}[ph]
  \begin{tabular}{l|l|c}
     \hline
     \multicolumn{2}{c|}{Type} & 95\% CL upper limit on $\Br(t \to qg)$ \\
     \hline
     \hline
     \multirow{4}{*}{$tug$} 
                            & BDT             & $6.73 \times 10^{-6}$ \\  
                            & BDT+$qg$-disc   & $6.38 \times 10^{-6}$ \\  
                            & \SaJa           & $5.61 \times 10^{-6}$ \\  
                            & \SaJa+$qg$-disc & $5.01 \times 10^{-6}$ \\  
     \hline
     \multirow{4}{*}{$tcg$} 
                            & BDT             & $5.89 \times 10^{-6}$ \\  
                            & BDT+$qg$-disc   & $5.69 \times 10^{-6}$ \\  
                            & \SaJa           & $4.40 \times 10^{-6}$ \\  
                            & \SaJa+$qg$-disc & $3.83 \times 10^{-6}$ \\  
     \hline
  \end{tabular}
  \caption{The 95\% CL upper limit on $\Br(t \to ug)$ and $\Br(t \to cg)$ by BDT without $qg$-discrimination variables and \SaJa\ with $qg$-discrimination variables}
  \label{table:UpperLimit}
\end{table}

\begin{figure}[ph]
  \subfloat[][ROC, $tug$ vertex]{
    \label{subfig:ROCOnlytug}
    \includegraphics[width=\plotsizeHalf]{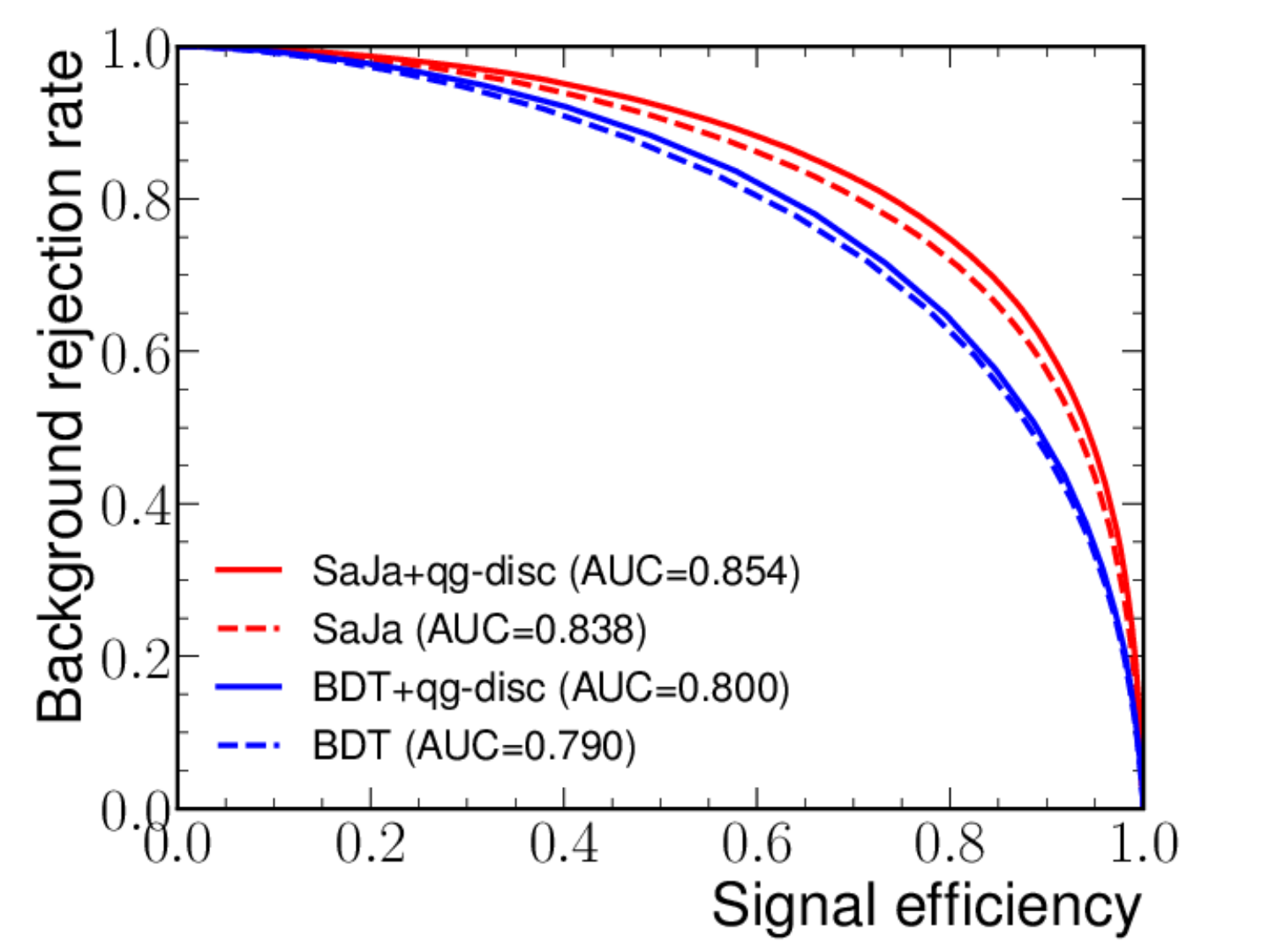}
  }
  \subfloat[][SIC, $tug$ vertex]{
    \label{subfig:SICOnlytug}
    \includegraphics[width=\plotsizeHalf]{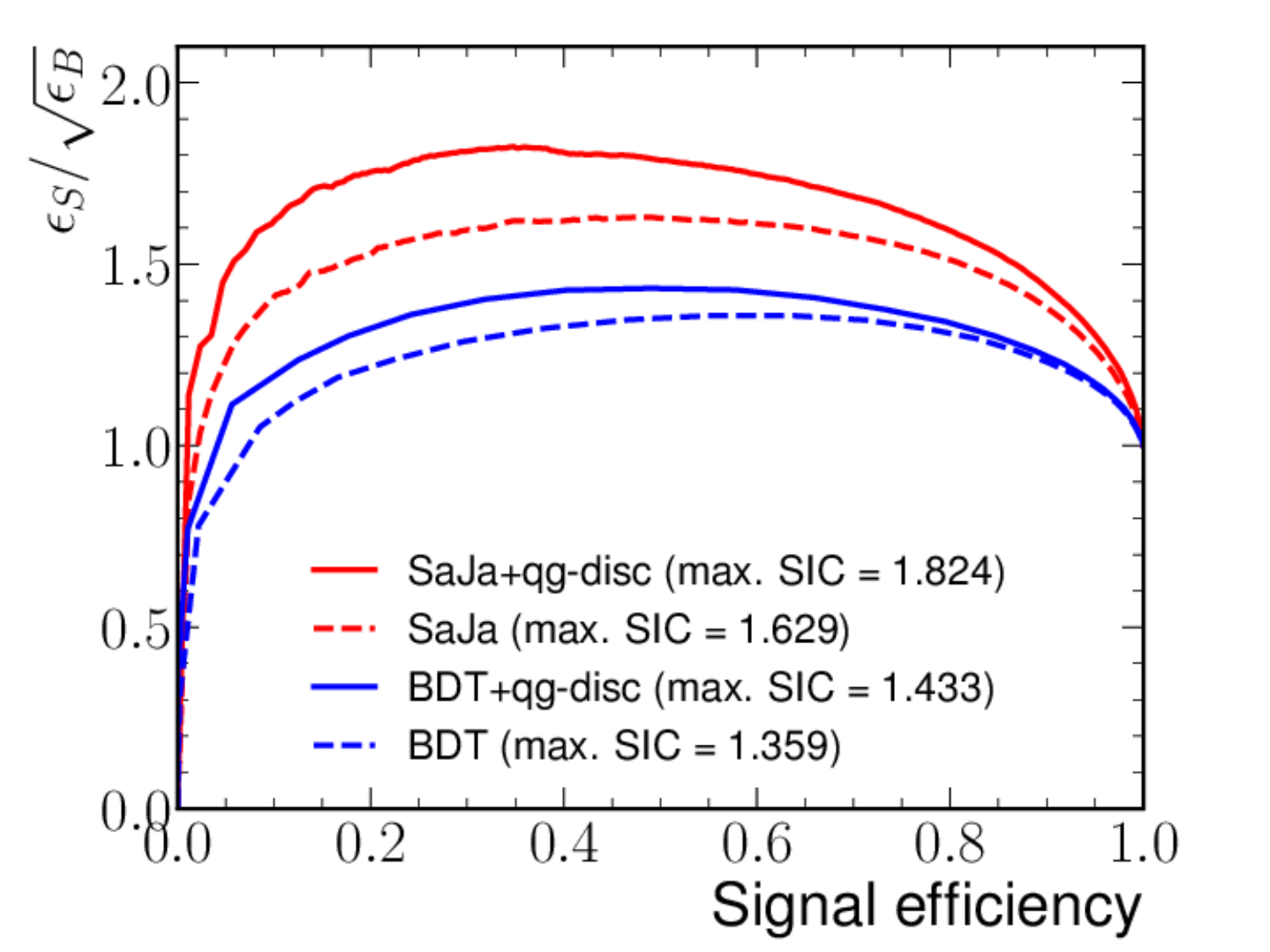}
  }
  
  \subfloat[][ROC, $tcg$ vertex]{
    \label{subfig:ROCOnlytcg}
    \includegraphics[width=\plotsizeHalf]{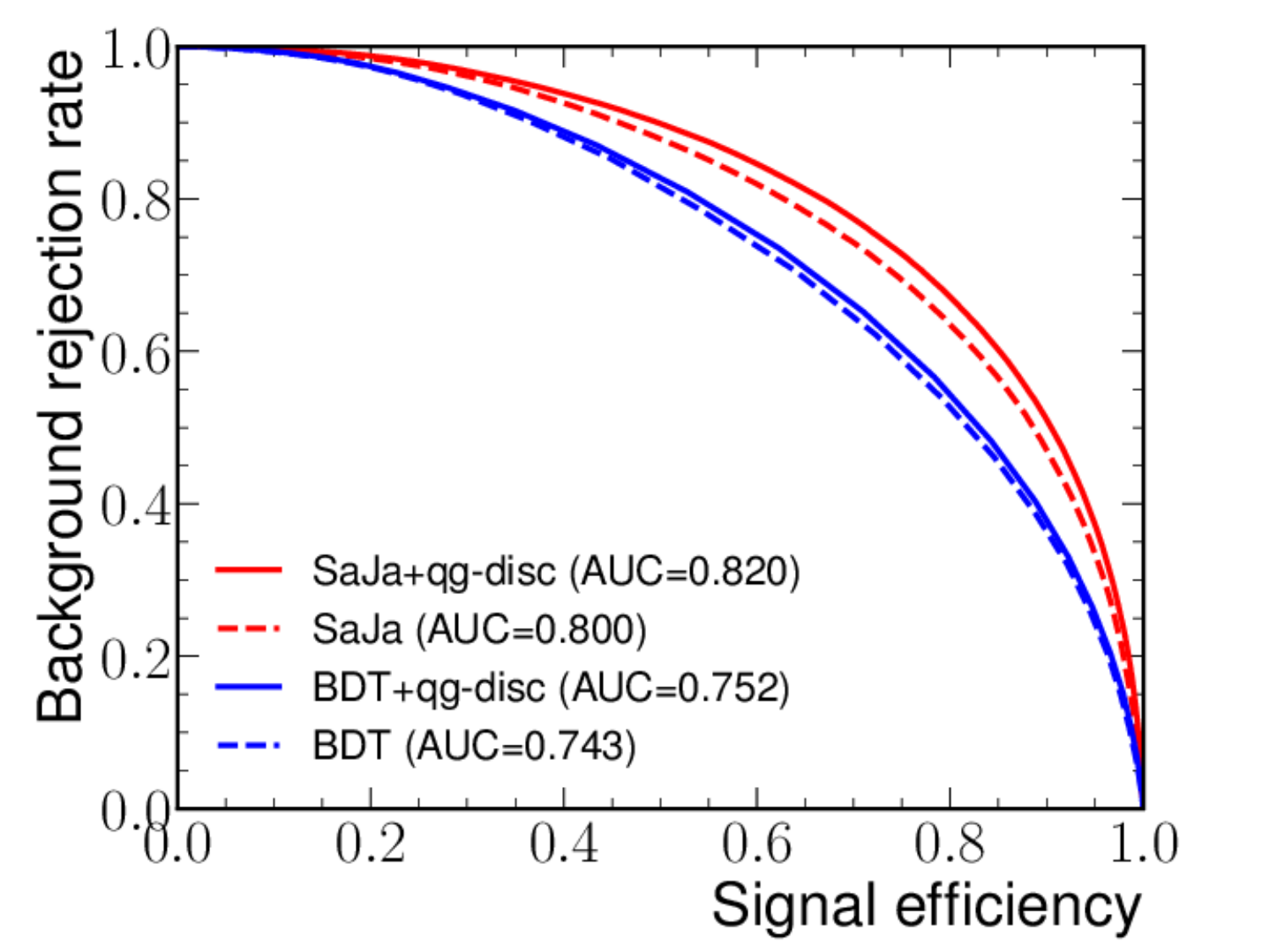}
  }
  \subfloat[][SIC, $tcg$ vertex]{
    \label{subfig:SICOnlytcg}
    \includegraphics[width=\plotsizeHalf]{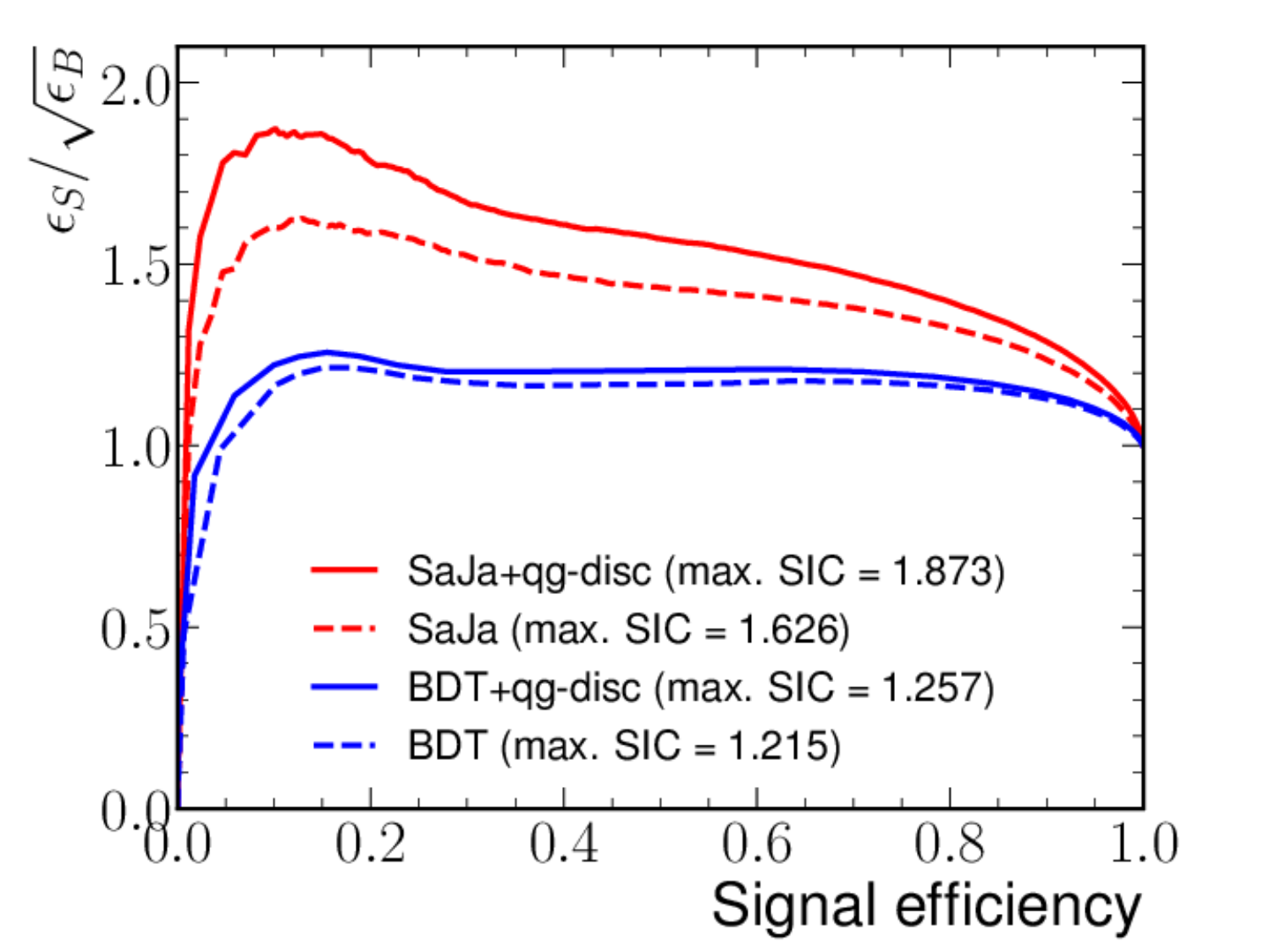}
  }
  \caption{
    The ROC curves (left column) and the SIC curves (right column) from BDT and \SaJa.
    The models are trained with signals having the $tug$ vertex (a, b) and the $tcg$ vertex (c, d).
    We use BDT (blue line) and \SaJa\ (red line) and train with $qg$-discrimination variable (solid line) and without $qg$-discrimination variable (dashed line).
    Max. SIC means the maximum value of $\epsilon_S / \sqrt{\epsilon_B}$.
  }
\label{fig:ROC}
\end{figure}

The ROC and SIC curves for each method, along with the values of AUC and maximum SIC, are shown in Figure~\ref{fig:ROC}.
The results are shown using signal samples originating from $tug$ vertex (\fig{subfig:ROCOnlytug} and \ref{subfig:SICOnlytug}) and $tcg$ vertex (\fig{subfig:ROCOnlytcg} and \ref{subfig:SICOnlytcg}) separately.
The blue dashed lines represent the baseline BDT performance. 
By including the $qg$-discrimination variables in the BDT, the blue solid line, we observe slight increases in both the AUC and the maximum SIC compared to the baseline. 
The \SaJa\ network without $qg$-discrimination variables, shown in the red dashed lines, significantly outperforms both BDTs for all signal efficiencies in both ROC and SIC.
Adding $qg$-discrimination variables to the \SaJa\ network, as shown in the red lines, improves the performance, like the BDT case, but the improvement for \SaJa\ is much more significant and results in the best overall performance.

We calculate the 95\% CL upper limits on $\Br(t \to qg)$ ($q = u, c$) to assess the impact on physics of our findings.
We choose events according to the threshold giving the maximum SIC for each model.
To find the expected upper limits, we assume the observation of the background-only hypothesis, with $B$ events, and find the number of signal events $S$, where the lower tail integral, up to $B$, of the Gaussian centered at $S+B$ with width $\sqrt{S+B}$ is 5\%, which is then translated into an upper limit on the signal cross section.
Utilizing the fact that the signal cross section is proportional to \cqgsqr, we derive the 95\% CL upper limits on \cqgsqr.
Finally, we determine the upper limits on $\Br(t \to qg)$ by employing the known relationship between \cqg and $\Br(t \to qg)$ \cite{PhysRevD91BranchingRatio}.
The expected upper limits from all methods are listed in \tbl{table:UpperLimit} assuming an integrated luminosity equivalent to the Run 2 of the LHC, 138 fb$^{-1}$.
The expected upper limits on $\Br(t \to qg)$ by \SaJa\ with $qg$-discrimination variables are 25\% and 35\% lower than the upper limits by the BDT baseline.

\section{Conclusion}
We investigate the effectiveness of using deep learning methods to enhance the sensitivity of collider searches for the $tqg$ FCNC process.
To enhance the distinction between the $tqg$ FCNC signals and SM backgrounds, we utilize $qg$-discrimination variables and introduce the \SaJa\ network to incorporate objects including all jets and the complete event topology.
Our study shows that the \SaJa\ network outperforms the BDT baseline, and with the addition of $qg$-discrimination variables, \SaJa\ makes further improvements. 
The performance is evaluated by the expected 95\% CL upper limits on the branching ratios $\Br(t \to ug)$ and $\Br(t \to cg)$, where the branching ratios obtained by \SaJa\ network with the $qg$-discrimination variables is 25\% and 35\% lower than those by BDT.

\newpage

\newpage

\newpage
\begin{acknowledgments}

This work was supported by the National Research Foundation of Korea (NRF) grant funded by the Korea government (MSIT). (No. 2021R1A2C1093704).
This research was supported by Basic Science Research Program through the National Research Foundation of Korea (NRF) funded by the Ministry of Education (2018R1A6A1A06024977).
This work was supported by the National Research Foundation of Korea (NRF) grant funded by the Korea government (MSIT).  (No. 2018R1C1B6005826).
This work was supported by the National Research Foundation of Korea (NRF) grant funded by the Korea government (MSIT).  (No. 2023R1A2C2002751).
This work was supported by the 2022 sabbatical year research grant of the University of Seoul.

\end{acknowledgments}

\end{document}